%% file: MSC.tex
\documentclass[twoside,12pt]{thesis}

\usepackage{floatflt,bm,psfrag,layout,enumitem}
\usepackage{amsmath}
\usepackage{lipsum}
\usepackage{graphicx,amssymb,rotate}
\usepackage{mathrsfs}
\usepackage{color}
\usepackage{graphicx}
\usepackage{subfigure}
\usepackage{multirow}
\usepackage{float}
\usepackage{pstricks}
\usepackage[toc,page]{appendix}
\usepackage{pifont}
\usepackage[utf8]{inputenc}
\usepackage{tabls}
\usepackage{makeidx}
\usepackage{wrapfig}
\usepackage{dashrule}
\usepackage[noadjust]{cite}

\newcommand{\quotes}[1]{``#1''}
\usepackage[bottom]{footmisc}
\usepackage[bookmarks=true,bookmarksnumbered=true,%
pdftitle={Thesis},%
pdfauthor={DEEPAK (IIT Indore)}]{hyperref}
\renewcommand\bibname{References}
\newcommand{\begp}{\begingroup}
 \newcommand{\eegp}{\endgroup}
\newcommand{\beq}{\begin{equation*}}
\newcommand{\eeq}{\end{equation*}}
\newcommand{\bea}{\begin{eqnarray}}
\newcommand{\eea}{\end{eqnarray}}

\def\nn{\nonumber}
\def\Eqn#1{Eq.\ (\ref{#1})}
\def\ra{\rightarrow}

\def\vector#1#2{\left( \begin{array}{c}#1\\ #2\end{array}\right)}

\def\order(#1){{\cal O} \left(#1 \right)}  
\def\Eqs#1#2{Eqs.\ (\ref{#1}) and (\ref{#2})}
\long\def\rpl#1!!#2!!{\textcolor{red}{#1} \textcolor{blue}{#2}} 

\def \cm{\cal{M}}

\def \ma{\mathcal{A}}
\def\m{\scriptstyle}

\newcommand{\nc}{\newcommand}

\def\nn{\nonumber\\}

\renewcommand{\Re}{{\rm Re \,}}
\renewcommand{\Im}{{\rm Im \,}}
\renewcommand{\bar}[1]{\overline{#1}}

\newcommand{\f}{\frac}
\newcommand{\baa}{\begin{array}}      \nc{\eaa}{\end{array}}
\newcommand{\bit}{\begin{itemize}}    \nc{\eit}{\end{itemize}}
\newcommand{\ben}{\begin{enumerate}}  \nc{\een}{\end{enumerate}}
\nc{\bce}{\begin{center}}     \nc{\ece}{\end{center}}
\nc{\bfl}{\begin{flushright}} \nc{\efl}{\end{flushright}}
\nc{\btb}{\begin{tabular}}    \nc{\etb}{\end{tabular}}
\def\a{\alpha}

\def\b{\beta}

\def\m{\mu}

\def\q2 {q^2}

\def\bt{\begin{table}}
\def\et{\end{table}}
\def\ocal{{\cal O}}
\def\lcal{{\cal L}}

\def\ie{ {\em i.e.,\ }}

\usepackage[compact]{titlesec}
\usepackage{setspace}
\titlespacing*{\section}{0pt}{3ex}{0ex}
\titlespacing*{\subsection}{0pt}{3ex}{0ex}
\titlespacing*{\subsubsection}{0pt}{3ex}{0ex}

\begin{document}



\include{FirstPage/FirstPage}

 \begin{singlespace}
{\hypersetup{linkbordercolor=white}
\tableofcontents
}
{
\hypersetup{linkbordercolor = white}
\listoffigures
}
\end{singlespace}
 \include{ABb/ABb}

\include{Chapter1/chapter1}

\include{Chapter2/chapter2}

\include{Chapter3/chapter3}

\include{Chapter4/chapter4}

\include{Chapter5/chapter5}

\appendix
\include{AppendixA/appendixA}
%

\include{Bibliography/bibliography}\end{document}

%% file: FirstPage/FirstPage.tex
\graphicspath{{FirstPage/}}
\pagestyle{plain}
 \pagenumbering{roman}\setcounter{page}{1}
\addcontentsline{toc}{section}{Title}

\begin{center}
\vspace*{-2 cm}
{\Large \textbf{Unitarity Bound on Extended Higgs  Sector with Higher Dimensional Operator}}\\

\vfill
{\large\textbf{M.Sc. Thesis}}\\
\vspace*{0cm}

\vfill\vspace*{0cm}
{\Large\textbf{by}}\\
\vspace*{0.5cm}
{\Large\textbf{DEEPAK}}\\
\vspace*{1cm}

\includegraphics[width=5cm,clip]{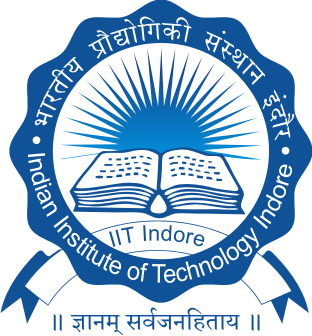}\\
\vspace*{2cm}

{\large\textbf{DISCIPLINE OF PHYSICS \\
    INDIAN INSTITUTE OF TECHNOLOGY INDORE \\
 JUNE, 2019}}\\
 
\vspace*{1cm}

\end{center}

\begin{center}
\newpage
\null
\newpage
\end{center}

\begingroup
\let\clearpage\relax
\let\cleardoublepage\relax
\let\cleardoublepage\relax

\begin{center}
\vspace*{-2.0cm}
{\Large \textbf{Unitarity Bound on Extended Higgs  Sector with Higher Dimensional Operator}}\\
\vspace*{0.3cm}
\vfill
{\large\textbf{A THESIS}}\\
\vspace*{0.3cm}

\vfill
{\large\it{Submitted in partial fulfillment of the \\requirements for the award of the degree}}\\
\vspace*{0.2cm}
{\large\textbf{$of$}}\\
\vspace*{0.5cm}

{\large\textbf{Master of Science}}\\
\vspace*{0.5cm}

\vfill\vspace*{0cm}
{\Large\textbf{by}}\\
\vspace*{0.7cm}
{\Large\textbf{DEEPAK}}\\
\vspace*{2cm}

\includegraphics[width=5cm,clip]{./Plots/IITimage}\\
\vspace*{2cm}

{\large\textbf{DISCIPLINE OF PHYSICS \\
    INDIAN INSTITUTE OF TECHNOLOGY INDORE \\
 JUNE, 2019}}\\
 
\vspace*{0.5cm}

\end{center}
\endgroup           
\newpage
\begin{center}
\begin{figure}
		\includegraphics[width=17.0cm]{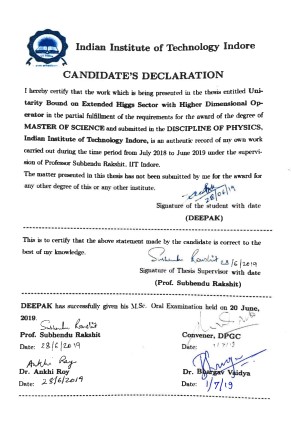}
	\end{figure}
\end{center}
 %


\setstretch{1.3} 

\newpage
\null
\newpage

\vspace*{5cm}
{\Huge\textbf{ \hspace{4cm} Dedicated}}\\
        
      {\Huge\textbf{  \hspace{5.8cm} to}}\\
 
  {\Huge\textbf{ \hspace{3.8cm} My Family}}
\null
\newpage
\newpage
\null
\newpage
\begin{center}
{\Huge\textbf{Acknowledgements}}\\

\phantomsection
\addcontentsline{toc}{section}{Acknowledgements}
\printindex

\end{center}
\vspace{28pt}
First and foremost, I gratefully acknowledge the continual guidance and support of my adviser, Prof. Subhendu Rakshit. His dedication to research and pursuit of
physics has been an invaluable source of inspiration and encouragement to me. His guidance helped me in all the time of research and writing of this thesis. 
\\I would like to thank my PSPC committee members, Dr. Ankhi Roy and Dr. Bhargav Vaidya for serving as my committee members even at hardship.
\\
It was not possible for me to achieve my M.Sc degree from an institute like IIT,
without the support of my family. I am very much thankful to my parents and
sisters for their constant support in my life.
\\
I would also like to thank all the research scholars working in Theoretical High
Energy Physics Lab for their co-operation and help during my M.Sc. research project.
\\
Special thanks to  Mr. Siddhartha Karmakar for valuable
guidance and support in thesis work.
\\


\newpage
\null
\newpage

\begin{center}
{\Huge\textbf{Abstract}}\\
\phantomsection
\addcontentsline{toc}{section}{Abstract}
\printindex
\end{center}
\vspace{28pt}


We study the perturbative unitarity bound given by higher dimensional bosonic operators up to dim-6 for the two-Higgs-doublet model (2HDM). They lead to new physics beyond the Standard Model.  We point out that such operators can lead to a larger cross-section in the vector boson fusion channels for the scalars, compared to the tree-level 2HDM.
We have obtained limits on a few bosonic operators up to dim-6 for 2HDM, by ensuring unitarity of the  S-matrix.

%% file: ABb/ABb.tex
\graphicspath{{ABb/}}

\newpage
{\Huge\textbf{List of Abbreviations}}\\
\pagestyle{plain}
\phantomsection
\addcontentsline{toc}{section}{List of abbreviations}
\printindex
\vspace{0pt}
\begp
\allowdisplaybreaks
\begin{align*}
\rm SM \hspace{3cm} & \rm Standard~Model\nn\\
\rm BSM \hspace{3cm} & \rm  Beyond~ Standard~Model  \nn\\
\rm NP \hspace{3cm} & \rm   New~Physics  \nn\\
\rm VEV \hspace{3cm} & \rm   Vacuum~expectation~value    \nn\\
\rm EW \hspace{3cm} & \rm   Electroweak   \nn\\
\rm EWSB \hspace{3cm} & \rm Electroweak~symmetry~breaking     \nn\\
\rm EWPT \hspace{3cm} & \rm Electroweak~precision~test  \nn\\
\rm LHC \hspace{3cm} & \rm  Large~Hadron~Collider    \nn\\
\rm  ATLAS \hspace{3cm} & \rm   A~Toroidal~LHC~Apparatus    \nn\\
\rm  CMS \hspace{3cm} & \rm   Compact~Muon~Solenoid    \nn\\
\rm 2HDM \hspace{3cm} & \rm  Two~Higgs~doublet~model     \nn\\
\rm SILH \hspace{3cm} & \rm  Strongly~interacting~ light~Higgs~model    \nn\\
\rm SMEFT \hspace{3cm} & \rm Standard~Model~Effective~Field~theory\nn\\
\rm SMEFT \hspace{3cm} & \rm Standard~Model~Effective~Field~theory\nn\\
\rm 2HDMEFT \hspace{3cm} & \rm Two~Higgs~doublet~Model~Effective~Field~theory\nn\\
\rm CM  \hspace{3cm} &  \rm  Center~of~Momentum \nn
\end{align*}
\eegp

%% file: Chapter1/chapter1.tex
\chapter{Introduction}
\label{chap:Intro}
\linespread{0.1}
\graphicspath{{Chapter1/}}
\pagestyle{headings}
\noindent\rule{15cm}{1.5pt} 
\pagenumbering{arabic}
From the centuries describing the constituents of matter and their interactions has been a goal of humankind. Already in the sixth century BC, the first ideas about
the smallest units forming larger structures arose in India. Around 450 BC Democritus coined the term ``atom” which is still in use today. It was not until 1967,
leaving out many significant milestones of discovery in nuclear and particle physics of
course, that our understanding of how to describe elementary particles drastically
improved. Combining electromagnetic and weak interaction incorporating the Higgs
mechanism, the Standard Model of particle physics was born. The Standard Model is one of the most successful theories which describes strong, weak, and electromagnetic forces and interactions between the elementary particles. The Lagrangian of SM has a particular type of mathematical symmetries due to which equations of motion derived from this Lagrangian have enabled physicists to make predictions about various observables which successfully tested in particle physics laboratories. The Higgs mechanism is essential to give rise to the masses of all the elementary particles through spontaneous symmetry breaking of the gauge symmetry. On July 4th 2012 at the Large Hadron Collider (LHC) discovery of a Higgs-like scalar boson confirms that the Higgs mechanism is responsible for electroweak symmetry breaking.
Even though the SM has been very successful in describing most of the elementary particles phenomenology, it is unable to explain various experimental observations, like neutrino mass, and the observed matter-antimatter asymmetry. Moreover, the Standard Model does not incorporate the theory of gravitation.
\section{ The Standard Model}
The standard model is the theory which is trying to
carry all fundamental forces of nature (except gravity) - under one umbrella and describe the nature of interactions between fundamental particles. These fundamental forces are strong, weak and electromagnetic interactions in terms of local gauge symmetries ${ SU(3)_C}$, ${ SU(2)_L}$, and ${ U(1)}_Y$ respectively. The local gauge group   ${ SU(3)_C}$  represents the strong interaction between quarks and gluons. Similarly,  group ${ SU(2)_L}$ is responsible for electroweak interactions.  The massive  $W^\pm$ and Z bosons, mediate the weak force whereas a massless vector gauge boson, the photon, mediate the electromagnetic force between electrically charged particles. There are three distinct gauge coupling constants, $g _1$, $g_2$, and $g_3$ corresponding to these groups, which help to determine the strength of the forces in SM.
The gauge coupling constant $g_3$ for group $SU(3)_C$ is large, e.g., $g_3=1.17$ at an  energy scale $M_t$ = 173 GeV.  
The electromagnetic force is small compared to the strong force due to the smallness of $g_1$ and $g_2$, e.g., $g_1 = 0.36$ and $g_2 = 0.67$ at the energy scale $M_t$.  These show that weak force is even smaller than the electromagnetic force and massive gauge boson mediator suppressed it.\\

According to the Standard  Model, the elementary particles classified into four categories:
\begin{itemize}
\item Quarks
\item Leptons
\item Gauge Bosons
\item Higgs Particle
\end{itemize}

\begin{figure}[ht]
\begin{center}
{
\includegraphics[width=4in,height=2.8in, angle=0]{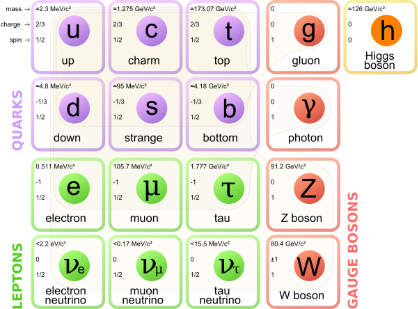}}
\caption{Fundamental particles make up the Standard Model of particle physics.}
\label{fig:SMparticle}
\end{center}
\end{figure}

Six quarks and six leptons and their antiparticles. Each quark comes with three colors namely red, green, and blue. The SM also includes gauge bosons such as photon, $W^\pm$, $Z$, and gluons and one neutral Higgs field (see Fig.~\ref{fig:SMparticle}). In the SM, the masses of all particles is given by Higgs particle only.
A left-handed $up$-$type$ quark and a left-handed $down$-$type$ quark together form a quark doublet under ${ SU(2)_L}$. Similarly, a left-handed charged lepton and the corresponding left-handed neutrino forms a doublet. Right-handed fermions are singlets under the same group.

The left-handed quark doublet $Q_L$ and lepton doublet $\chi_L$ are denoted as,
\begingroup
\allowdisplaybreaks \beq
Q_L \equiv \begin{pmatrix}
u_{L} \\  d_{L}
\end{pmatrix} \,
\qquad {\rm and} \qquad
\chi_L \equiv \begin{pmatrix}
\nu_{lL} \\  l_{L}
\end{pmatrix}  ,
\label{lepquarlL}
\eeq
\endgroup\par
Right-handed quark singlet $q_R$ and lepton singlet $l_R$ are given by,
\begingroup
\allowdisplaybreaks 
\bea
 q_R=u_{R},~d_{R};~~&{\rm and }&~~ l_{R},\label{lepquarlR}
\eea
\endgroup\par
where $u$ represents the $up$-$type$ quarks of the three generations $u,c,t$; and $d$ stands for the $down$-$type$ quarks $d,s,b$. The charged leptons are represented  by $l = e,\mu,\tau$ with the corresponding left-handed neutrinos $\nu_l = \nu_e,\nu_\mu,\nu_\tau$. SM does not deal with right-handed neutrinos.
Also, the complex doublet scalar  field in the SM  is given by,
\begp
\allowdisplaybreaks \beq
\varphi= \begin{pmatrix}
\varphi^+ \\  \varphi^0
\end{pmatrix}\nn,
\eeq
\eegp\par
where $\varphi^+=\frac{\varphi_1+i\varphi_2}{\sqrt{2}}$ is the complex charged scalar field and $\varphi^0=\frac{\varphi_3+ i \varphi_4}{\sqrt{2}}$ is the neutral complex scalar field. These 
complex scalar transform as a doublet under ${ SU(2)_L}$. 

 The SM Lagrangian is given by,
\begingroup
\allowdisplaybreaks
\beq
\mathcal{L}_{SM} = \mathcal{L}_{fermions} + \mathcal{L}_{gauge} + \mathcal{L}_{Higgs} + \mathcal{L}_{Yukawa}.\nn
\eeq
\endgroup\par
The kinetic terms of the fermions and their interactions with gauge bosons is given by,  
\begingroup
\allowdisplaybreaks \beq
\mathcal{L}_{fermions}= i \bar{Q}_L~\gamma^{\mu} D^L_\mu~ Q_L+i \bar{\chi}_L~\gamma^{\mu} D^L_\mu ~\chi_L  + i\bar{q}_R~\gamma^{\mu} D^R_\mu~ q_R+ i \bar{l}_R~\gamma^{\mu}D^R_\mu ~l_R,
\label{fermionew}
\eeq
\endgroup\par
where the covariant derivative of fermion doublet with left chirality is defined as,
\begingroup
\allowdisplaybreaks \beq
D^L_\mu=\left(\partial_{\mu}+i g_2 T^a W^a_{\mu}+i g_1 \frac{Y}{2} B_{\mu} \right),
\label{coVderi}
\eeq
\endgroup\par
and the covariant derivative of singlet fermion with right chirality is given by,
\begingroup
\allowdisplaybreaks \beq
D^R_\mu=\left(\partial_{\mu}+i g_1 \frac{Y}{2} B_{\mu}\right). 
\eeq
\endgroup\par
The second and third terms of the equation~\ref{coVderi} are related to ${ SU(2)_L}$ and ${ U(1)}_Y$ gauge transformations respectively. $W^{a}_\mu$ ($a$=1,2,3) are the ${ SU(2)_L}$ gauge bosons, corresponding to three generators of ${ SU(2)_L}$ group and $B_\mu$ is the ${ U(1)}_Y$ gauge boson. In the SM, the generators of ${ SU(2)_L}$ are $2\times2$ matrices $ T^a=\frac{1}{2} \tau^a$, where the $\tau^a$, are the Pauli spin matrices,
\begingroup
\allowdisplaybreaks \beq
\tau^1= \begin{pmatrix}
0 & 1\\ 1& 0
\end{pmatrix},
\qquad
\tau^2= \begin{pmatrix}
0 & -i\\ i& 0
\end{pmatrix},
\qquad
\tau^3= \begin{pmatrix}
1 & 0\\ 0& -1
\end{pmatrix}.\nn
\eeq
\endgroup\par
$Y$ is the weak hypercharge operator, generator of ${ U(1)}_Y$ group. The hypercharge operator, defined as a linear combination of the electromagnetic charge operator $Q$ and the third generator $T^3=\frac{\tau^3}{2}$ of ${ SU(2)_L}$, is given by,
\begingroup
\allowdisplaybreaks \beq
Y=2(Q-T^3). \nn
\eeq
\endgroup\par

The gauge part of the Lagrangian contains the kinetic term and interaction term of the gauge bosons.
It is given by,
\begingroup
\allowdisplaybreaks
\beq
\mathcal{L}_{gauge}= -\frac{1}{4} {W}_{\mu\nu}^a { W}^{a,\mu\nu}- \frac{1}{4} B_{\mu\nu} B^{\mu\nu}.\nn
\eeq
\endgroup\par
The field strength tensors are defined as, 

\begingroup
\allowdisplaybreaks
\beq
{B}_{\mu\nu}=\partial_\mu B_\nu - \partial_\nu B_\mu ,\nn
\eeq
\endgroup\par

\begingroup
\allowdisplaybreaks
\beq
{ W}_{\mu\nu}^a=\partial_\mu W_\nu^a - \partial_\nu W_\mu^a - g_2 \epsilon^{abc} W_{\mu}^b W_{\nu}^c\nn, 
\eeq
\endgroup\par

where $\epsilon^{abc}$ is structure constant of $SU(2)_L$ group such that $[T^a,T^b]=i\epsilon^{abc} T^c$.

In the SM, the gauge symmetry prevents us from adding explicit mass terms for gauge bosons. As a result, in the limit of exact symmetry, all gauge bosons are massless.
To incorporate the massive $W^\pm$ and $Z$ bosons into the SM, the Higgs mechanism has been developed to circumvent this constraint on the mass. In this mechanism, the masses of all particles (except neutrinos) are obtained through the spontaneous breaking of the $SU(2)_L \otimes U(1)_Y$ gauge symmetry at the electroweak scale.

The part of the Lagrangian, which is responsible for the masses of the gauge bosons and the Higgs and also to the interaction between the Higgs and the gauge bosons, is given by,
\begingroup
\allowdisplaybreaks \bea
\mathcal{L}_{Higgs} =& (D^{L,\mu} \varphi)^\dagger (D^L_\mu \varphi) - V(\varphi),
\label{LagHiggs}
\eea
\endgroup\par
where $V(\varphi)$ is the SM Higgs potential, and is given by,
\begingroup
\allowdisplaybreaks \bea
 V(\varphi) &=& m^2 \varphi^\dagger \varphi + \lambda (\varphi^\dagger \varphi)^2,\label{SMSSBpot}\\
 {\rm with,}~~~
 \varphi &\equiv& \vector{\varphi^+}{\varphi^0}= \vector{\frac{\varphi_1+i\varphi_2}{\sqrt{2}}}{\frac{\varphi_3+ i \varphi_4}{\sqrt{2}}},\label{PotScalefield}
\eea
\endgroup\par
The electroweak symmetry breaking and how the particles get masses will be discussed in the following.

Robert Brout,  Fran\c{c}ois Englert, and  Peter Higgs group proposed the electroweak symmetry breaking (EWSB) which is also known as  Higgs mechanism in  Standard Model~\cite{Higgs:1964pj,Higgs:1966ev}. In this mechanism, the real component $\varphi_3$ of the neutral complex scalar of the electroweak doublet acquires a non-vanishing vacuum expectation value (VEV) leading to EWSB. As a result, the gauge group ${SU(2)_L \otimes U(1)_Y}$ is broken down to $U(1)_{EM}$, the symmetry group that corresponds to electromagnetism.

In the SM, the Higgs potential $V(\varphi)$, which is responsible for spontaneous symmetry breaking, is given in eqn.~\ref{SMSSBpot},  
\begingroup
\allowdisplaybreaks 
\beq
V(\varphi) = m^2 \varphi^\dagger \varphi + \lambda (\varphi^\dagger \varphi)^2\nn.
\eeq
\eegp\par
For $\lambda<0$, the potential goes to $-\infty$ as $|\varphi|\rightarrow \infty$, i.e., it gets unbounded from below at very high field values. So $\lambda$ is taken to be positive. If $m^2>0$, the minimum of the potential is found at $|\varphi|=\sqrt{\left\langle 0| \varphi^\dagger \varphi |0\right\rangle} = 0$, where $|0\rangle$ represents the ground state. However, the minimum occurs at $|\varphi|=\sqrt{\left\langle 0| \varphi^\dagger \varphi |0\right\rangle} = \sqrt{-\frac{m^2}{2\lambda}}=\frac{v}{\sqrt{2}}$ for $m^2<0$ and $\lambda>0$. In the former case, the symmetry is unbroken while in the latter case symmetry is apparently broken.
\begin{figure}[ht]
\begin{center}
{
\includegraphics[width=2.8in,height=2.in, angle=0]{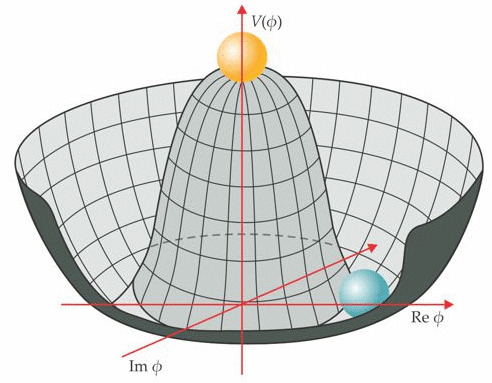}}
\caption{Schematic diagram of the Higgs potential. The potential is symmetric about the vertical axis and broken at minimum. The circular line indicates the remaining U(1) symmetry.}
\label{fig:SMPotSSB}
\end{center}
\end{figure}

The fields $\varphi_1$, $\varphi_2$ and $\varphi_4$ in eqn.~\ref{PotScalefield} are not physical fields and do not receive any VEV. They can be removed by a particular gauge choice, known as unitary gauge choice. These unphysical fields, known as the Goldstone bosons, are `eaten' by the massless $W^\pm$ and the $Z$ bosons, so that they get massive. In this gauge, the scalar field can be written as,
\begingroup
\allowdisplaybreaks \beq
\varphi= \begin{pmatrix}
0 \\  \frac{h+v}{\sqrt{2}}
\end{pmatrix} 
,
\label{SSBphi2}
\eeq
\endgroup\par
where $\varphi_3=h+v$, $h$ is the physical Higgs boson. Using the eqns.~\ref{SMSSBpot}, and~\ref{SSBphi2}, the eqn.~\ref{LagHiggs} can be written explicitly as,
\begingroup
\allowdisplaybreaks
 \begin{align}
\mathcal{L}_{Higgs} = & \frac{1}{2} \partial_\mu h \partial^\mu h + (h+v)^2 \left( \frac{g_2^2}{4} W_{\mu}^+ W^{\mu-} + \frac{g_1^2 + g_2^2}{8} Z_\mu Z^\mu  \right) \nn
& -\frac{1}{4} \lambda (h+v)^2 ( (h + v)^2 - 2  v^2) + ...
\label{MassWZh}
\end{align}
\endgroup\par
The charged $W^\pm$ bosons are defined as $W_{\mu}^\pm = \frac{W_{\mu}^1 \mp i W_{\mu}^2}{\sqrt{2}}$. The $Z$ boson and photon are orthogonal combinations of $W_\mu^3$ and $B_\mu$: $Z_\mu= c_W W_\mu^3 - s_W B_\mu$ and $A_\mu= s_W W_\mu^3 + c_W B_\mu$.
$c_W \equiv \cos\theta_W$ and $s_W \equiv \sin\theta_W$, where $\theta_W$ is the Weinberg angle. It can be expressed in terms of the gauge coupling constants as,
\begingroup
\allowdisplaybreaks \beq
\theta_W = \cos^{-1}\left(\frac{g_2}{\sqrt{g_1^2 + g_2^2}}\right).
\label{ThetaWein}
\eeq
\endgroup\par
One can express the electric charge in terms of the gauge coupling constants $g_1$ and $g_2$ as, $e=\frac{g_1 g_2}{\sqrt{g_1^2 +g_2^2}}$, which determines the strength of the electromagnetic interaction. 

The mass terms for $W^\pm$ and $Z$ bosons as well as for the Higgs boson $h$ from eqn.~\ref{MassWZh} can be identified as,
\begin{equation}
M_W^2 = \frac{1}{4} g_2^2 v^2,
\end{equation}

\begin{equation}
M_Z^2 =\frac{1}{4} (g_1^2 + g_2^2) v^2,
\end{equation}

\begin{equation}
M_h^2 = 2 \lambda v^2,
\end{equation}
where the photon $A_\mu$ remains massless after symmetry breaking, i.e., the vacuum breaks the original symmetry  $ SU(2)_L \times U(1)_Y$ in such a way that only the ${U(1)_{EM}}$ survives as the residual symmetry with a conserved charge $Q=T_3+\frac{Y}{2}$.
\section{Organization of the Thesis}
 The organization of the thesis is as follows:

This chapter starts general introduction to Standard Model theory. 
In the second Chapter, Extended scalar sector such as type-II two Higgs doublet model have been reviewed. After that, effective field theory approach for heavier NP Particles  in terms of higher dimensional operator discussed.
The  expression  of the longitudinal vector gauge boson scattering of the processes like 
(a)$ W_L^{\pm} W_L^{\mp}\rightarrow W_L^{\pm} W_L^{\mp}$, (b) $W_L^{\pm} W_L^{\pm}\rightarrow W_L^{\pm}W_L^{\pm}$ and (c)$ W_L^{+} W_L^{-} \rightarrow Z_L Z_L $ have been reviewed, and these amplitudes are changed on the inclusion of the dimension-six operator and it has been discussed in the third Chapter. 
In the fourth Chapter, the unitarity of scattering matrix and unitarity bound  given by higher dimensional bosonic operators up to dim-6 for two-Higgs-doublet model (2HDM) has been discussed.\\
The last Chapter provides the summary and the conclusions of this thesis work.

%% file: Chapter2/chapter2.tex
\chapter{Extended Higgs Sector}
\label{chap:Extended}
\linespread{0.1}
\graphicspath{{Chapter2/}}
\pagestyle{headings}
\noindent\rule{15cm}{1.5pt}
\section{Introduction}
In 2012, the collaborations  of  ATLAS and CMS at the Large Hadron Collider (LHC) at the CERN discovered a new scalar resonance with a mass $\sim 125$ GeV which the last missing piece of the Standard Model (SM) – the Higgs boson \cite{Chatrchyan:2012xdj,Aad:2012tfa}. From experimental data, the measured value of the Higgs mass, $m_h\approx125$ GeV, it shows that the Higgs potential of the SM is not stable up to very high energy scales.
Although, the discovered Higgs boson may be SM Higgs or just one of the Higgs in the extended SM. One well-motivated direction for discovering physics beyond the SM is to search
for additional Higgs bosons. 
The SM relies on the minimal choice of a single SU(2) scalar doublet giving masses to all the massive particles contained in the SM. Extension of the SM scalar sector is a common practice in constructing BSM models.
\newline
Adding a second SU(2) doublet scalar to the SM field content represents one of the simple possible extensions of the Standard Model (SM) of strong and electroweak (EW) interactions. Apart from the direct search for bounds on the parameter space of 2HDM ~\cite{Akeroyd:2000wc}.
They are various types of theoretical constraints imposed on 2HDM like perturbativity, unitarity and vacuum stability, etc. For the 2HDM to behave as a perturbative quantum field theory at any given scale, one must impose the conditions like $ | \lambda_i |  \leq 4 \pi$ $( i = 1,2, . . , 6)$. On applying such conditions, one implies upper bounds on the values of the couplings at low as well as high scales.\\

If beyond 2HDM any new physics exists as a decoupled sector from the mass scale of 2HDM, the outcome of such new physics can be expressed in terms of the higher-dimensional operators in an effective theory where the fields of 2HDM constitute the low-energy spectrum~\cite{Crivellin:2016ihg}. Such an effective theory is dubbed as the two-Higgs-doublet model effective field theory (2HDMEFT) in the literature. A complete basis of the Operators upto dimension-six in 2HDMEFT has been introduced only recently~\cite{Karmakar:2017yek}. We impose a more stringent condition of unitarity on 2HDM tree-level with the 6-dim operators, and we evaluate bounds on higher dimensional bosonic operators up to dim-6.

\section{Two Higgs Doublet Model}
 
In the 2HDM, we introduce two ${\rm SU}(2)_L$ doublets $\varphi_{i}$ ($i=1,2$)\cite{Akeroyd:2000wc}:
\begin{eqnarray}
\varphi_i =\frac{1}{\sqrt{2}} \begin{pmatrix} \sqrt{2} w_i^+ \\ (h_i+v_i)
  +iz_i \end{pmatrix} \,.
\end{eqnarray}
where $v_i$ are the vacuum expectation values (VEV) of the neutral components, satisfying $v_1^2+v_2^2= v^2$, with $v=$ 246 {\rm GeV}. The 
ratio of VEVs is defined as $\tan\beta\equiv v_2/v_1$. The 2HDM Lagrangian for $\varphi_i$ can be written as

\begin{equation}
 \mathcal{L} =   \sum_i |D_\mu \varphi_i|^2 - V(\varphi_1,\varphi_2) + \mathcal{L}_{\mathrm{Yuk}}, 
 \label{eq:lagrangian}
\end{equation}
where the first term denotes the kinetic term for the two Higgs doublets, $V(\varphi_1, \varphi_2)$ is the Higgs potential and the 
last term denotes the Yukawa interactions between $\varphi_i$ and the SM fermions.
\subsection{The Scalar Potential}
There are two equivalent notations that are used in the literature~\cite{Bhattacharyya:2015nca} to
write the 2HDM scalar potential with a softly broken $Z_2$ symmetry
($\varphi_1\to \varphi_1$, $\varphi_2\to -\varphi_2$)~:
\paragraph*{\em Parametrization 1 ~:}
\begin{eqnarray}
V(\varphi_1,\varphi_2) &=& m_{11}^2 \varphi_1^\dagger\varphi_1 +
m_{22}^2\varphi_2^\dagger\varphi_2 -\left(m_{12}^2 \varphi_1^\dagger\varphi_2
+{\rm h.c.} \right) +\frac{\beta_1}{2} \left(\varphi_1^\dagger\varphi_1
\right)^2 \nonumber \\ && +\frac{\beta_2}{2} \left(\varphi_2^\dagger\varphi_2 \right)^2
 +\beta_3 \left(\varphi_1^\dagger\varphi_1 \right)
\left(\varphi_2^\dagger\varphi_2 \right) +\beta_4 \left(\varphi_1^\dagger\varphi_2
\right) \left(\varphi_2^\dagger\varphi_1 \right) \nonumber \\ &&+\left\{\frac{\beta_5}{2}
\left(\varphi_1^\dagger\varphi_2 \right)^2 +{\rm h.c.} \right\} \,.
\label{notation1}
\end{eqnarray}
\paragraph*{\em Parametrization 2 ~:}
\begin{eqnarray}
 V &=& \lambda_1 \left(\varphi_1^\dagger\varphi_1 - \frac{v_1^2}{2}
 \right)^2 +\lambda_2 \left(\varphi_2^\dagger\varphi_2 - \frac{v_2^2}{2}
 \right)^2 +\lambda_3 \left(\varphi_1^\dagger\varphi_1 +
 \varphi_2^{\dagger}\varphi_2 - \frac{v_1^2+v_2^2}{2} \right)^2 \nonumber
 \\ && +\lambda_4 \left( (\varphi_1^{\dagger}\varphi_1)
 (\varphi_2^{\dagger}\varphi_2) - (\varphi_1^{\dagger}\varphi_2)
 (\varphi_2^{\dagger}\varphi_1) \right) + \lambda_5 \left({\rm Re}~
 \varphi_1^\dagger\varphi_2 - \frac{v_1v_2}{2} \right)^2 \nonumber
 \\ &&+ \lambda_6
 \left({\rm Im}~ \varphi_1^\dagger\varphi_2 \right)^2 \,.
\label{notation2}
\end{eqnarray}
\\
The bilinear terms proportional to $m_{12}^2$ in equation(2.2.3)
  or $\lambda_5$ in equation(2.2.4) 
 breaks the $Z_2$ symmetry softly. 
cast in the form of equation(2.2.3). 
The connections between the parameters of equation(2.2.3)
 and equation(2.2.4) 
  are given below~\cite{Bhattacharyya:2015nca}:
\begin{eqnarray}
&& m_{11}^2 =-(\lambda_1v_1^2+\lambda_3v^2)~;~ m_{22}^2=
  -(\lambda_2v_2^2+\lambda_3v^2)~;~
  m_{12}^2=\frac{\lambda_5}{2}v_1v_2~;~
  \beta_1=2(\lambda_1+\lambda_3)~; \nonumber \\ &&
  \beta_2=2(\lambda_2+\lambda_3)~;~ \beta_3=2\lambda_3+\lambda_4~;~
  \beta_4=\frac{\lambda_5+\lambda_6}{2}-\lambda_4~;~
  \beta_5=\frac{\lambda_5-\lambda_6}{2}~.
\end{eqnarray}
\\
In equation(2.2.5), 
 $v=\sqrt{v_1^2+v_2^2} = 246$ GeV, where $v_1$
and $v_2$ are the vevs of the two doublets $\varphi_1$ and $\varphi_2$
respectively. For the part of this article, we choose to work with the notation of Eq. (2.2.4).\newline
After EW symmetry breaking, the physical 2HDM scalar spectrum consists of five states: two CP-even Higgses $h$, $H$ with $m_h< m_{H},$ 
a CP-odd scalar $A$ and a charged scalar pair $H^\pm,$ which may be written as
\begin{eqnarray}
\left(\begin{array}{c}
H\\
h
\end{array}\right) = 
\left(
\begin{array}{cc}
c_{\alpha}& s_{\alpha}\\
-s_{\alpha} & c_{\alpha}
\end{array}
\right)
\,\,
\left(\begin{array}{c}
h_1\\
h_2
\end{array}\right), \quad\quad \quad \quad \quad \quad \quad\nonumber \\
\\
\left(\begin{array}{c}
G\\
A
\end{array}\right) = 
\left(
\begin{array}{cc}
c_{\beta}& s_{\beta}\\
-s_{\beta} & c_{\beta}
\end{array}
\right)
\,\,
\left(\begin{array}{c}
z_1\\
z_2
\end{array}\right)\,\,,\quad  
\left(\begin{array}{c}
G^{\pm}\\
H^{\pm}
\end{array}\right) = 
\left(
\begin{array}{cc}
c_{\beta}& s_{\beta}\\
-s_{\beta} & c_{\beta}
\end{array}
\right)
\,\,
\left(\begin{array}{c}
w^{\pm}_1\\
w^{\pm}_2
\end{array}\right), \nonumber
\end{eqnarray}\\
where, $c_{\beta(\alpha)} \equiv \cos\beta(\alpha)$ and $s_{\beta(\alpha)} \equiv \sin\beta(\alpha)$.\\
The mixing angle of the CP-even sector is defined through the following relation:

\begin{eqnarray}
\tan 2\alpha =\frac{2\left(\lambda_3+\frac{\lambda_5}{4}
  \right)v_1v_2}{\lambda_1v_1^2-
  \lambda_2v_2^2+\left(\lambda_3+\frac{\lambda_5}{4}\right)(v_1^2-v_2^2)}
\,.
\label{angle_alpha}
\end{eqnarray}
\\
The Goldstone bosons $G$ and $G^\pm$ are absorbed as longitudinal components of the $Z$ and $W^\pm$ bosons. In the limit $c_{\beta-\alpha} = 0$ 
(the {\it alignment limit} for $h$), the state $h$ can be identified with the SM Higgs, 
its couplings to fermions and gauge bosons being precisely those predicted by the SM.
Note that, equation (2.2.3) and (2.2.4), both contain eight free parameters. In the notation
of equation(2.2.4), these are $v_1$, $v_2$ and six $\beta_i$ couplings. We can trade $v_1$ and $v_2$
for $v = \sqrt{v_1^2 + v_2^2}$ and $\tan\beta$. Except for $\lambda_5$, all other $\lambda$ parameters may be traded
for four physical scalar masses ($m_h,m_H,m_A$ and $m_{H^+}$) and the angle, $\alpha$. The equivalence 
of these two sets of parameters is demonstrated by the following relations:
\begin{subequations}
\begin{eqnarray}
\lambda_1 &=& \frac{1}{2v^2\cos^2\beta}\left[m_H^2\cos^2\alpha
  +m_h^2\sin^2\alpha
  -\frac{\sin\alpha\cos\alpha}{\tan\beta}\left(m_H^2-m_h^2\right)\right]
\nonumber
 \\ &&-\frac{\lambda_5}{4}\left(\tan^2\beta-1\right) \,, \\
\lambda_2 &=& \frac{1}{2v^2\sin^2\beta}\left[m_h^2\cos^2\alpha
  +m_H^2\sin^2\alpha
  -\sin\alpha\cos\alpha\tan\beta\left(m_H^2-m_h^2\right) \right]
\nonumber
 \\ &&-\frac{\lambda_5}{4}\left(\cot^2\beta-1\right) \,, \\
\lambda_3 &=& \frac{1}{2v^2}
\frac{\sin\alpha\cos\alpha}{\sin\beta\cos\beta}
\left(m_H^2-m_h^2\right) -\frac{\lambda_5}{4} \,, \\
\lambda_4 &=& \frac{2}{v^2} m_{H^+}^2 \,, \\
\lambda_6 &=& \frac{2}{v^2} m_A^2 \,.
\end{eqnarray}
\label{inv2HDM}
\end{subequations}
\\
Among these, $v$ is already known to be 246 GeV and it is likely that the
lightest CP-even Higgs is what has been observed at the LHC, then
$m_h$ is also known (125 GeV). The rest of the parameters need to be
constrained from theoretical as well as experimental considerations.

\section{Higher Dimensional Operator}
In general, at  higher energies 2HDM theory is  replaced by  
some theory of new physics (NP) which need to  satisfy the following \cite{Crivellin:2016ihg} :
\begin{itemize}
\item[({\it i})] Its gauge group contains the SM gauge group~$SU(3)_C\times SU(2)_L\times U(1)_Y$ as a subgroup.
\item[({\it ii})] It contains two Higgs doublets as dynamical degrees of freedom.
\item[({\it iii})] At low energies it reproduces the 2HDM.
\end{itemize}
The new physics theory which satisfies above condition have heavier NP particles which are  integrated out and their outcome are  characterized in terms of Wilson coefficients of higher-dimension operators overcome by inverse powers of $\Lambda$. In our 2HDM case we have
\begin{eqnarray}
\mathcal{L}_{2HDM} = \mathcal{L}_{2HDM}^{(4)}   + \f{1}{\Lambda }  \sum_{i} C_{i}^{(5)} O_{i}^{(5)} + \f{1}{\Lambda^2} \sum_{i} C_{i}^{(6)} O_{i}^{(6)}  + \ocal\!\left(\f{1}{\Lambda^3}\right)\,.
\end{eqnarray} 
\label{eqn:Leff} 
Here $\lcal_{\mathrm 2HDM}^{(4)}$ is the standard renormalizable 2HDM Lagrangian. $O_{i}^{(5)}$ generalize the Weinberg operator~\cite{Weinberg:1979sa} and $O_i^{(6)}$ denote the dimension-six operators. $C_i^{(5)}$ and $C_i^{(6)}$ are their dimensionless Wilson coefficients.
We neglect the effects of operators of dimension-seven and higher, which are overcome by at least three powers of $\Lambda$ and also not considering the effect of the operator of dimension-five. In this thesis, we only interested in the operator of dimension-six.
\subsection{Operator Basis}
 According to Universal theories~\cite{Rattazzi:2004ph}, the deviations in the Higgs boson properties from SM depicted only in terms of higher dimensional bosonic operators. Both the Warsaw basis and Strongly interacting light Higgs(SILH) basis are bosonic bases,\ie all bosonic operators are kept in those bases ~\cite{Karmakar:2017yek}. 
We include all the dimension six operators in our basis of 2HDMEFT, which is inspired by the Strongly interacting light Higgs(SILH) basis of SMEFT. The total Lagrangian along with dimension-six operators  looks like~\cite{Karmakar:2017yek}:

\begin{equation}
\mathcal{L} = \mathcal{L}_{2HDM}^{(4)}  + \mathcal{L}^{(6)},
\end{equation}
where,
\begin{eqnarray}
\mathcal{L}^{(6)} &=& \mathcal{L}_{\varphi^4 D^2} + \mathcal{L}_{\varphi^2 D^2 X} + \mathcal{L}_{\varphi^2 X^2} + \mathcal{L}_{\varphi^6} + \mathcal{L}_{\varphi^3 \psi^2} + \mathcal{L}_{\varphi^2 \psi^2 D} + \mathcal{L}_{\varphi \psi^2 X} + \mathcal{L}_{D^2 X^2} + \mathcal{L}_{\psi^4}.\nn
\end{eqnarray}

We have defined our notation as follows: $\varphi$, $\psi$ and $X$ stand for the two scalar doublets, fermions and gauge field strength tensors respectively. $D$ stands for a derivative. Throughout this thesis, we have worked under the definition of $\mathcal{L} \supset c_{i} ( O_{i} /\Lambda^2 )$, which means all the Wilson coefficients are named according to the suffix of the corresponding operator. For example, $c_{Bjk}$ is the Wilson coefficient of $O_{Bjk}$. In this thesis, we have worked under bosonic operator only.
\begin{itemize}
\item $\varphi^6$: Operators with Higgs doublets only, which modify the Higgs potential. These are the corrections to the potential of the renormalizable 2HDM~\cite{Karmakar:2017yek}.
\item  $\varphi^4 D^2$: Operators with four Higgs doublets and two derivatives,
 which recast the kinetic terms of the Higgs fields, the Higgs-gauge boson interactions and the W and Z masses.
 \item $\varphi^2 X^2$: Operators with two Higgs doublets and  two field strength tensors.
 \item $\varphi^2 D^2 X $: Operators  with  two Higgs doublets, two derivatives two field strength tensors.  These operators contribute to observables, applicable to precision tests and SM-like Higgs phenomenology.
\end{itemize}
The bounds on Wilson coefficients are around  $\mathcal{O}(10^{-3})$ for $\varphi^2 X^2$ and $\varphi^2 D^2 X^2$ types operator, which are  directly given by measurement of the decay width  $h \rightarrow \gamma \gamma, Z \gamma$~\cite{Pomarol:2013zra}. Due to which these operators are irrelevant for our purpose.

%% file: Chapter3/chapter3.tex
\chapter{Vector Boson Scattering with Six-dimensional operators}
\label{chap:VV Scattering 2hdmeft}
\linespread{0.1}
\graphicspath{{Chapter3/}}
\pagestyle{headings}
\noindent\rule{15cm}{1.5pt} 
\section{Introduction}
Additional scalars in 2HDM that couple with the W and Z bosons, longitudinal vector boson scattering along with scalar exchanges should provide a compatible way to direct search methods to probe into the scalar sector. The unitarity of the S-matrix for the longitudinal electroweak vector boson scattering $ V_L V_L \rightarrow V_L V_L $ preserved by the Higgs boson in the SM.
In the SM, the Higgs boson helps preserve the unitarity of the S-matrix for the longitudinal
electroweak vector boson scattering $ V_L V_L \rightarrow V_L V_L $.
 The Higgs boson mediated diagram precisely cancels the residual $s$-dependence (where $\sqrt{s} $ denotes the energy in the centre-of-mass frame), thus taming the high energy behaviour of the cross-section appropriately. With an extended scalar sector, the preservation of unitarity could be a more complex process. In this section, we discuss about the restoration of unitarity with six-dimensional operator i.e., $\phi^4 D^2$ operator. To the best of our knowledge, these works  have not yet been presented in the literature.

\section{Vector Boson Scattering under $\varphi^4 D^2$ Operator}
One can express  $ V_L V_L \rightarrow V_L V_L $
scattering  amplitude as
\begin{equation}
    \mathcal{M} = A_4 E_{cm}^4 + A_2 E_{cm}^2 + A_0 + A_{-2} E_{cm}^{-2} +....
\end{equation}
where, $E_{cm}$ is the centre-of-mass energy.
 We need the scalar particles  in the model so $ A_2 $  becomes exactly zero for $E_{cm}>> M_i(i \equiv W, Z, h, H )$
and theory becomes unitarized, i.e., the cross-section will decrease with energy.
The gauge and scalar contributions to $A_2$ and $A_0 $ are denoted as,
\begin{equation*}
  A_2 = A_{2,g} + \sum_{S} A_{2,S}
\end{equation*}
where,
$ S=h,H $.
\\By Ref. \cite{Khan:2016mdq}, the expressions for $A_{2}$ in 2HDM are given by :
\\
1. $W_L^+ W_L^- \rightarrow W_L^+ W_L^-$
\begin{equation}
      A_2=  \frac{g_2^2(4 M_W^2-3 c_W^2 M_Z^2)(1+x)}{2  M_W^4}  -\frac{g_2^2}{2 M_W^2}~C^2~(1+x)
 \end{equation}
 2. $W_L^+ W_L^+ \rightarrow W_L^+ W_L^+$
 \begin{equation}
    A_2 =  \frac{g_2^2(3 c_W^2 M_Z^2-4 M_W^2)}{ M_W^4} + \frac{g_2^2}{M_W^2}~C^2 
 \end{equation}
 3. $W_L^+ W_L^- \rightarrow Z_L Z_L$
 \begin{equation}
   A_2 =  \frac{g_2^2 c_W^2 M_Z^2}{ M_W^4}  -\frac{g_2^2}{c_W M_W M_Z}~C C^{\prime} 
 \end{equation}
 where, $M_V$ is the mass of $V(\equiv W^\pm, Z)$, coupling multipliers $ C = \cos(\beta-\alpha),$  $C' = \sin(\beta-\alpha)$, $x \equiv \cos\theta$, $\theta$ is the scattering angle.
 
 The $\varphi^4 D^2$ operators reformulate  the Higgs fields \cite{Karmakar:2018scg}, giving  rise to the rescaling of the $hVV$ couplings.
 
 \begin{equation}
     \cos(\beta-\alpha) \rightarrow \cos(\beta-\alpha)(1-x_{2})+ \sin(\beta-\alpha) y
      \end{equation}
 \begin{equation}
    \sin(\beta-\alpha) \rightarrow  \sin(\beta-\alpha)(1-x_{1})+ \cos(\beta-\alpha) y
 \end{equation}
 where, $x_1$, $x_2$ and $y$ are functions of the Wilson coefficients of the higher dimensional operators and are given by:
 \begin{eqnarray*}
 x_1 &=&\frac{v^2}{f^2} \Big( c_{H1} c_{\b}^2  s_{\a}^2 + c_{H2} c_{\a}^2 s_{\b}^2 + \frac{1}{8} c_{H1H2} s_{2\a} s_{2\b} + c_{H12} (c_{\a}^2 c_{\b}^2 + s_{\a}^2 s_{\b}^2 - \frac{1}{4} s_{2\a} s_{2\b})\\
&&\hspace{50pt}+ c_{H1H12} c_{\b} s_{\a} (s_{\a} s_{\b} - \frac{1}{2} c_{\a} c_{\b}) + c_{H2H12} c_{\a} s_{\b} (c_{\a} c_{\b} - \frac{1}{2} s_{\a} s_{\b}) \Big) ,\\
x_2 &=&\frac{v^2}{f^2} \Big( c_{H1} c_{\b}^2 c_{\a}^2 + c_{H2} s_{\a}^2 s_{\b}^2 + \frac{1}{8} c_{H1H2} s_{2\a} s_{2\b} + c_{H12} (s_{\a}^2 c_{\b}^2 + c_{\a}^2 s_{\b}^2 - \frac{1}{4} s_{2\a} s_{2\b})\\
&&\hspace{50pt}+ c_{H1H12} c_{\b} c_{\a} (c_{\a} s_{\b} - \frac{1}{2} s_{\a} c_{\b}) + c_{H2H12} s_{\a} s_{\b} (s_{\a} c_{\b} - \frac{1}{2} c_{\a} s_{\b}) \Big),\\
y &=& \frac{v^2}{f^2} \Big( \frac{1}{2} c_{H1} s_{2\a} c_{\b}^2 -\frac{1}{2} c_{H2}  s_{2\a} s_{\b}^2 - \frac{1}{8} c_{H1H2} c_{2\a} s_{2\b} - \frac{1}{2} c_{H12} (  c_{2\b}  s_{2\a} + \frac{1}{2} c_{2\a}  s_{2\b}) \\
&&\hspace{50pt}+ \frac{1}{4} c_{H1H12} ( s_{2\a} s_{2\b} - c_{2\a} c_{\b}^2 ) - \frac{1}{4} c_{H2H12} ( s_{2\a} s_{2\b} + c_{2\a} s_{\b}^2 )\Big).
\end{eqnarray*}
As a result of reformulating of fields, couplings of both the CP-even neutral scalars to vector bosons and fermions get altered compared to 2HDM at the tree-level.
With the $\varphi^4 D^2$ operator, we found an expression for $A_2$ in terms of $\cos({\beta-\alpha})$ and $ \tan{\beta}$ for various $ V_L V_L $ scattering process.
In the figs.(1), (2) and (3) show the plot between $A_2$ and $\cos({\beta-\alpha})$ at fixed value of $ \tan{\beta}$ for various $ V_L V_L $ scattering process in 2HDM tree level and 2HDM with $\varphi^4 D^2$ operator.

\begin{figure}[ht]
    \begin{center}
    \includegraphics[height=15em]{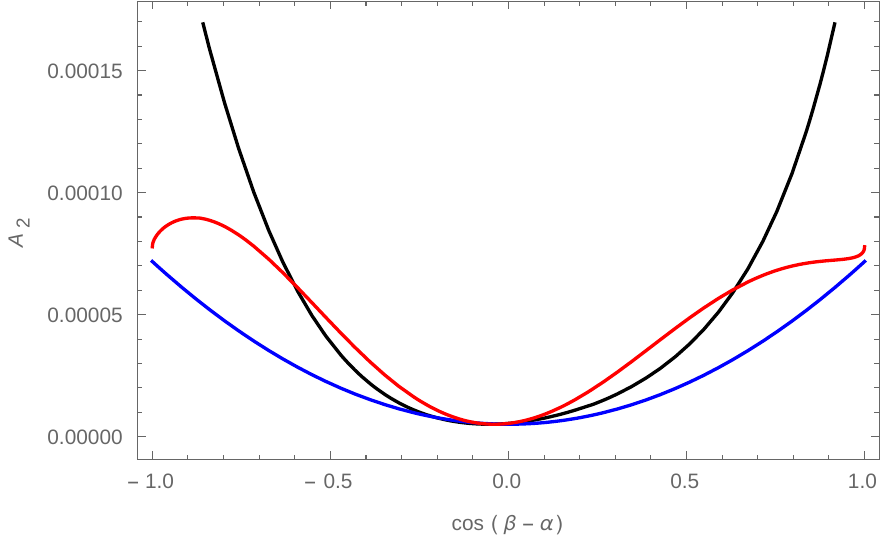}
    \caption{The effect of $\varphi^4 D^2$ on $A_2$ for $ W_L^{\pm} W_L^{\mp}\rightarrow W_L^{\pm} W_L^{\mp}$. The black line shows the variation of $A_2$ with
$\cos({\beta-\alpha})$ for 2HDM. Blue and red lines show the  variation of $A_2$ for $ \tan{\beta}$ equal to 1 and 5 respectively in 2HDM with the 6-dim. operator.
    }
    \label{fig:my_label}
    \end{center}
\end{figure}
\newpage
\begin{figure}[ht]
    \begin{center}
    \includegraphics[height=15em]{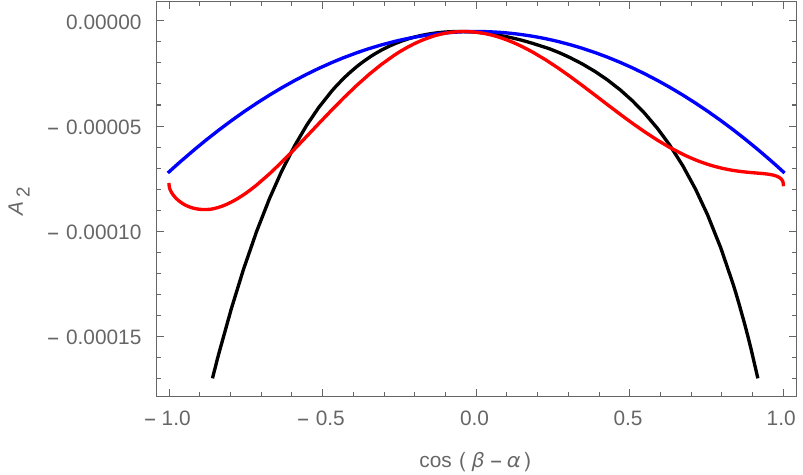}
    \caption{The effect of $\varphi^4 D^2$ on $A_2$ for $W_L^{\pm} W_L^{\pm}\rightarrow W_L^{\pm}W_L^{\pm}$. Colour coding is the same as in fig.1.}
    \label{fig:my_label}
    \end{center}
\end{figure}
\begin{figure}[h!]
    \begin{center}
    \includegraphics[height=15em]{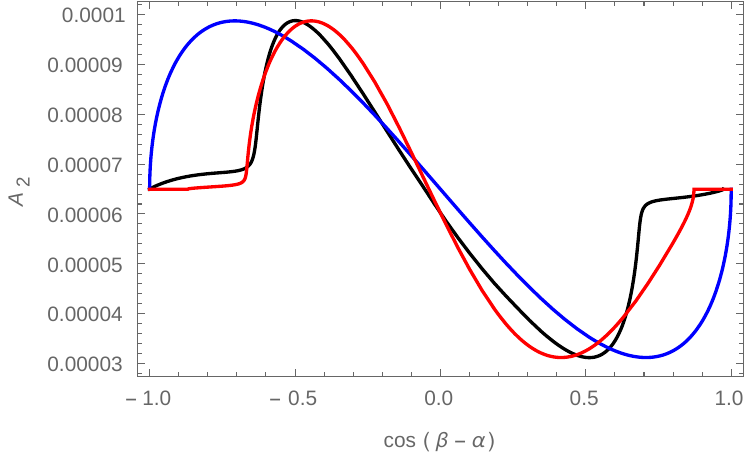}
    \caption{The effect of $\varphi^4 D^2$ on $A_2$ for $ W_L^{+} W_L^{-} \rightarrow Z_L Z_L $.  Colour coding is the same as in fig.1.}
    \label{fig:my_label}
    \end{center}
\end{figure}
The change in $A_2$ in presence of the  six-dimension  terms shows that the cross-sections of these processes also change in 2HDMEFT compared to 2HDM. This can affect the discovery potential of the new scalar.


%% file: Chapter4/chapter4.tex
\chapter{Unitarity of the Scattering Matrix}
\label{chap:Unitarity of the scattering matrix}
\linespread{0.1}
\graphicspath{{Chapter4/}}
\pagestyle{headings}
\noindent\rule{15cm}{1.5pt} 
In general, scattering matrix also called S-matrix is an asymptotic operator which describes how particles going into a scattering event transform into particles going out. The S-matrix can be determined using the Hamiltonian description; it does not require a Hamiltonian or Lagrangian description of the intermediate details of the scattering at all, it can be built up without regard to the local space-time structure. We do not need any knowledge of the local structure of space and time to talk about incoming and outgoing particles since these are defined at far away locations and far away times. The elements of S-matrix  are scattering amplitude for the particular scattering process and it is the probability amplitude of the outgoing spherical wave relative to the incoming plane wave in a stationary-state scattering process in quantum physics.
We know that every scattering amplitude can be expanded in terms of the partial waves as follows:
\begin{eqnarray*}
 {\cal M}(\theta) = 16 \pi \sum_{l=0}^{\infty}(2l+1) a_l  P_l(\cos\theta) \,,
\label{uni_amp}
\end{eqnarray*}
where, $\theta$ is the angle of scattering. 
Every partial wave amplitude is bounded from the `unitarity' condition:

\begin{eqnarray*}
|a_l| \le 1 \,.
\end{eqnarray*}

\quotes{In quantum physics, Unitarity is equivalent to the conservation of probability. A violation of unitarity is identical to a violation of the principles of quantum mechanics—this is too sacred a principle to give up!}~\cite{H.E. Haber}.
In this chapter, we will discuss the unitarity constraint on the dimension-six operator for 2HDM for that we will calculate the S-matrix, which is mentioned in the appendix. To the best of our knowledge, these work have not yet been presented in the literature.
\section{ Quantum Mechanical Approach}
\begin{figure}
\centering
\includegraphics[scale=0.6]{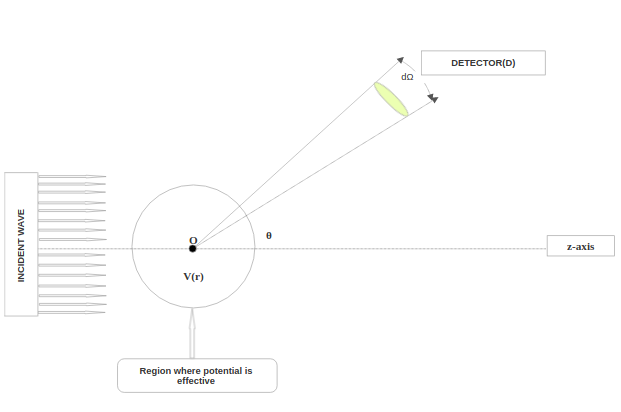}
\caption[Scattering in QM]{\em Schematic diagram for scattering by a potential $V(\vec{r})$.}
\label{qm-scat}
\end{figure}

Let us consider a scattering experiment in which a steady incident  beam is maintained for an indefinitely long time, 
i.e. the incident flux, $F_{\rm in}$, is constant. Then, there will be a steady stream of scattered particles too.
In figure ~\ref{qm-scat}, the incident mono-energetic beam is parallel to the $z$-axis and is assumed to be much wider than the zone of influence of 
the potential, $V(\vec{r})$, centred at O. Far from this zone of influence a detector,
 D, measures the number, $dn$, of particles scattered per unit time into the solid angle $d\Omega$, 
centred around the direction defined by the polar angles $\theta$ and $\phi$.
The number, $dn$, is proportional to $F_{\rm in}$ and to $d\Omega$; the constant of proportionality, $d\sigma/d\Omega$, is defined to be the differential scattering
 cross-section in the direction ($\theta$,$\phi$). Thus
 \begin{equation}
     dn=F_{\rm in}\cdot d\Omega \cdot\frac{d\sigma}{d\Omega} \,.
 \label{cross section}
 \end{equation}
 In the quantum theory of scattering, we imagine that an incident plane wave, $\psi_{\rm in}= Ae^{ikz}$, traveling along the z-axis with normalization A. The wave  encounters a scattering potential producing an outgoing spherical wave. At large distances from the 
scattering center, one can decompose the wave function$\psi(\vec{r})$ into a part $e^{ikz}$ describing the incident beam and another part $\psi_{\rm sc}$   for the scattered particles. As the collision is elastic, i.e., the energies of the incident and scattered particle are the same.Thus we may write,
\begin{equation}
\psi(\vec{r})_{_{r\to\infty}} \approx e^{ikz}+\underbrace{f(\theta,\phi)\frac{e^{ikr}}{r}}_{\psi_{\rm sc}} \,.
\label{basic}
\end{equation}
In this expression, only the function $f(\theta,\phi$), which is called the scattering amplitude, depends on the interaction of the incident particles.For momentum dependent coupling between particles, the function $f$ is expressed as $f\equiv f(k,\theta, \phi)$ to include dependence on the incident energy and momentum of particles. 
The $\phi$ dependence should be included in the general case, to account for the anisotropy of the potential. 
However, if the target is azimuthally symmetrical, the $\phi$ dependence would no longer be present. Note that, the 
spherical wave carries a factor of  $1/r$, because this portion of $|\psi|^2$ is spherically diverging and must go like  $1/r^2$  to conserve probability.
\subsection{Differential Cross-section}
We know  that the expression for the current density  $J(\vec{r})$ associated with a wave function $\psi(\vec{r})$ is:
\begin{equation}
\vec{J}(\vec{r})=\frac{\hbar}{2im}\left[\psi^*(\vec{\nabla}\psi)-(\vec{\nabla}\psi^*)\psi\right]=\frac{1}{m}{\rm Re}\left[\psi^*(\vec{r})\frac{\hbar}{i}\vec{\nabla}\psi(\vec{r})\right] \,.
\end{equation}
The incident and scattered fluxes are obviously proportional to the normal components of $\vec{J}_{\rm in}$ and $\vec{J}_{\rm sc}$ respectively. We will call the proportionality constant $C$.
Since $\psi_{\rm in} = e^{ikz}$, we obtain
\begin{eqnarray}
&& \left(J_{\rm in}\right)_z = \frac{1}{m}{\rm Re}\left[e^{-ikz}\frac{\hbar}{i}\frac{\partial}{\partial{z}}e^{ikz}\right]=\frac{\hbar k}{m}\,, \\
&\Rightarrow& F_{\rm in} = C \left(J_{\rm in}\right)_z \,.
\end{eqnarray}
For radially diverging scattered wave, the number of particles crossing an area $d\vec{s}=ds \,\hat{r}$, subtending solid angle $d\Omega$ at the origin is
\begin{eqnarray}
dn=\underbrace{C\vec{J}_{\rm sc}}_{\vec{F}_{\rm sc}}.(ds\hat{r})=C(J_{\rm sc})_rds \,.
\end{eqnarray}
Clearly, it is the $r$-th component of $\vec{J}_{sc}$ which receives our attention. Remembering $\psi_{\rm sc}=\frac{1}{r}f(\theta,\phi)e^{ikr}$ we may get,
\begin{eqnarray}
(J_{\rm sc})_r=\frac{1}{m} {\rm Re} \left[f^*(\theta,\phi)\frac{e^{-ikr}}{r}\frac{\hbar}{i}\frac{\partial}{\partial{r}}\{f(\theta,\phi)\frac{e^{ikr}}{r}\}\right]
=\frac{1}{m}|f(\theta,\phi)|^2\frac{\hbar{k}}{r^2}
\end{eqnarray}
Hence,
\begin{eqnarray}
dn=\frac{C}{m}|f(\theta,\phi)|^2\hbar{k}\frac{ds}{r^2}=F_{\rm in}|f(\theta,\phi)|^2d\Omega \,.
\end{eqnarray}
Comparing this with \Eqn{cross section} we obtain,
\begin{eqnarray}
\frac{d\sigma}{d\Omega}=|f(\theta,\phi)|^2 \,.
\end{eqnarray}
Now, the problem of determining the scattering cross-section reduces to finding the scattering amplitude, $f(\theta,\phi)$, in quantum mechanics. The quantity, $f(\theta,\phi)$, actually gives  us 
information  about the `probability amplitude' for
scattering in a direction $(\theta,\phi)$, and hence is related to the differential cross-section which is the quantity of interest for the experimentalists. The scattering amplitude  is obtained by solving the Schr\"{o}dinger equation under the scattering potential. Depending on the mathematical form of the potential, there are several methods to find the scattering amplitude. The method of partial waves, in particular, comes in handy when the potential is central.
\subsection{Method of Partial Waves}
In the special case of a central potential $V(r)$, the orbital angular momentum $\vec{L}$ of the particle is a constant
 of motion. Therefore, there exists stationary states with well defined angular momentum, {\it i.e.}, eigenstates common to $H$, 
$L^2$ and $L_z$. We shall call such wave functions `partial waves' and denote them as
$\psi_{klm}(\vec{r})$. Their angular dependence is always given by the spherical harmonics $Y_l^m(\theta,\phi)$ --
the potential $V(r)$ influences their radial parts only.
We know that $e^{ikz}$ is a solution of the Schr\"{o}dinger equation with $V(r)=0$ in the $\{H,~p_x,~p_y,~p_z\}$ basis and may be denoted 
by $|0,0,k \rangle $ where z-axis is chosen as the direction of motion. Now if we wish, we may translate our wave function in 
terms of $\psi_{klm}(\vec{r}) \equiv$ $R_{kl}(r)Y_l^m(\theta,\phi)$ which are the eigenfunctions in the $\{H,~L^2,~L_z\}$ basis. For a free particle we know that $R_{kl}(r)$ is a linear combination of spherical Bessel and Neumann functions. 
But as Neumann function blows up at the origin it is dropped out. So we may write,
\begin{eqnarray}
\psi_{klm}^{(0)}(r,\theta,\phi)=j_l(kr)Y_l^m(\theta,\phi) \,,
\end{eqnarray}
where, the superscript `0' denotes that these are `free' (the potential is identically zero) spherical waves. Let us 
connect these two sets of bases as follows:
\begin{equation}
e^{ikz}=\sum_{l=0}^{\infty}\sum_{m=-l}^{+l}{\ma}_l^m(k)j_l(kr)Y_l^m(\theta,\phi) \,,
\end{equation}
where ${\ma}_l^m(k)$ are suitable expansion coefficients that can only be functions of the magnitude of the momentum. Since the LHS of above equation is independent of $\phi$, we require that RHS should also be independent of $\phi$, {\it i.e.} $m=0$. Thus, we are left with,
\begin{eqnarray}
e^{ikz}=\sum_{l=0}^{\infty}{\ma}_l^0(k)(k)j_l(kr)Y_l^0(\theta) \,,
\label{eikz}
\end{eqnarray}
where $Y_l^0(\theta)$ is given by
\begin{eqnarray}
Y_l^0(\theta)=\sqrt{\frac{(2l+1)}{4\pi}}P_l(\cos\theta) \,.
\end{eqnarray}
In view of this, we introduce the following shorthand:
\begin{eqnarray}
{\ma}_l=\sqrt{\frac{(2l+1)}{4\pi}}{\ma}_l^0 \,.
\end{eqnarray}
Using this, one may rewrite \Eqn{eikz} as
\begin{eqnarray}
e^{ikz}=\sum_{l=0}^{\infty}{\ma}_l(k)j_l(kr)P_l(\cos\theta) \,.
\label{eikz1}
\end{eqnarray}
To determine ${\ma}_l(k)$, we need to use the following integral representation for the Bessel function:
\begin{eqnarray}
j_l(kr)=\frac{1}{2i^l}\int_{-1}^{+1}P_l(\cos\theta)e^{ikr\cos\theta}d(\cos\theta) \,.
\end{eqnarray}
To illustrate the use of the above formula, let us multiply \Eqn{eikz1} by $P_{l'}(\cos\theta)d(\cos\theta)$ and integrate between $-1$ to $+1$ to obtain
\begin{eqnarray}
&& \int_{-1}^{+1}P_{l'}(\cos\theta)e^{ikr\cos\theta}d(\cos\theta) = \frac{2}{(2l'+1)}\sum\limits_{l=0}^{\infty}{\ma}_l(k) j_l(kr) \delta_{ll'} \,, \nonumber \\
&\Rightarrow& 2i^{l'}j_{l'}(kr) = \frac{2}{(2l'+1)} {\ma}_{l'}(k) j_{l'}(kr) \,, \nonumber \\
&\Rightarrow& {\ma}_{l'}(k) = i^{l'}(2l'+1) \,.
\end{eqnarray}
Plugging this into \Eqn{eikz1} we get the final expression as
\begin{eqnarray}
e^{ikz}=\sum_{l=0}^{\infty}i^l(2l+1)j_l(kr)P_l(\cos\theta) \,.
\label{eikz2}
\end{eqnarray}
In view of the asymptotic form of the Bessel function,
\begin{eqnarray}
j_l(kr)\xrightarrow{r\to\infty} \frac{\sin(kr-\frac{l\pi}{2})}{kr} \,,
\end{eqnarray}
we may rewrite \Eqn{eikz2} as
\begin{equation}
e^{ikz} \xrightarrow{r\to\infty} \frac{1}{2ikr}\sum_{l=0}^{\infty}(2l+1)\left[e^{ikr}-e^{-ikr}(-1)^l \right] P_l(\cos\theta) \,.
\label{asym1}
\end{equation}
Thus, at large distances, each $\psi_{klm}^{(0)}$ and so the `whole' $e^{ikz}$ results from the superposition of a converging spherical wave, $e^{-ikr}/r$, and a diverging spherical wave, $e^{ikr}/r$, whose amplitudes differ only by a phase. The fact that the squared amplitudes for both the incoming and outgoing spherical waves are same, simply reflects the conservation of probability. The presence of a scattering potential can only affect the amplitude of the outgoing spherical wave. Since probability conservation demands that the magnitude of the amplitude for the diverging wave should not change, it can only pick up additional phases arising due to the presence of a scattering potential. We will see the details in the following subsection.
\subsubsection*{Presence of a central potential -- asymptotic modification of the radial part}
\begin{wrapfigure}{r}{0.4\textwidth}
\centering
 \includegraphics[scale=0.5]{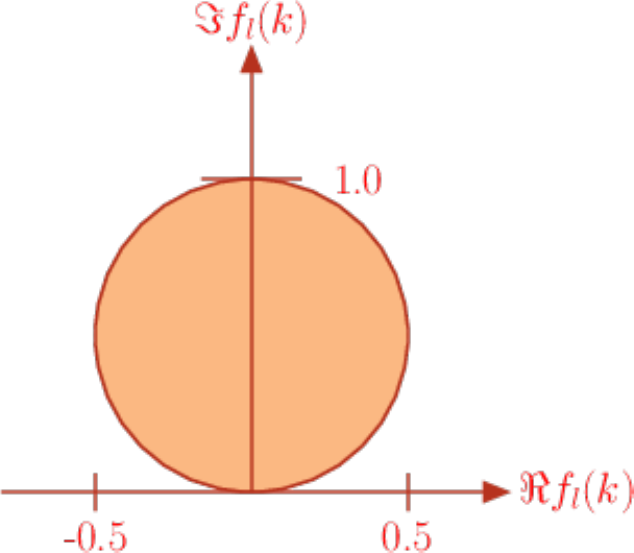}
 \caption{Unitarity circle.}
 \label{unicirc}
\end{wrapfigure}

The previous subsection was devoted for $V(r)=0$. Presence of a central potential simply modifies the wave function from the plane wave nature. But we know a special thing -- whatever be the form of $V(r)$, it dies out at a finite distance and in the asymptotic limit we should get the wave function in the form of Eq.~(\ref{basic}). 
In the presence of $V(r)$ the radial part of Schrodinger equation reads:
\begin{equation}
\left[\frac{d^2}{dr^2}+\frac{2}{r}\frac{d}{dr}+\left\{ k^2-\frac{2m}{\hbar^2}V(r)-\frac{l(l+1)}{r^2}\right\} \right] R_{kl}(r)=0 \,.
\label{tise}
\end{equation}
We assume that the potential is short-ranged, {\it i.e.}, $V(r)\rightarrow{0}$ as $r\rightarrow\infty$. Then, at large distances, Eq.~(\ref{tise}) reduces to the free-particle equation. Therefore, the solution of Eq.~(\ref{tise}) should asymptotically approach the general solution for free particle:
\begin{equation}
R_{kl}(r)\xrightarrow{r\to\infty} A_lj_l(kr)+B_l\eta_l(kr)=\frac{C_l}{kr}\sin(kr-\frac{l\pi}{2}+\delta_l) \,,
\label{mod}
\end{equation}
where, the last step follows from the asymptotic forms of Bessel and Neumann functions:
\begin{eqnarray}
j_l(kr)\xrightarrow{r\to\infty}\frac{\sin(kr-\frac{l\pi}{2})}{kr} \,, ~~~~ \eta_l(kr)\xrightarrow{r\to\infty}-\frac{\cos(kr-\frac{l\pi}{2})}{kr} \,.
\end{eqnarray}
The quantities, $C_l$ and $\delta_l$ are related to $A_l$ and $B_l$  as follows:
\begin{eqnarray}
\tan\delta_l=-\frac{B_l}{A_l}\,, ~~~ {\rm and}~~ C_l=\sqrt{A_l^2+B_l^2} \,.
\end{eqnarray}
Note that, unlike the free-particle case, here we did not demand $B_l=0$. This is due to the lack of information about 
the potential, we do not know the actual behavior of the wave function near the origin. \Eqn{mod} only represents the radial wave function at large distances where the potential is ineffective. Thus, the total wave function far away from the scatterer can be written as
\begin{eqnarray}
&& \psi(\vec{r})_{r\rightarrow\infty}=\sum_{l=0}^{\infty}R_{kl}(r \to \infty)P_l(\cos\theta) \,, \\
&\Rightarrow& \psi(\vec{r})_{r\rightarrow\infty}=\frac{1}{2ikr}\sum_{l=0}^{\infty}C_le^{-i\delta_l}e^{-i\frac{l\pi}{2}}
\left[e^{ikr}e^{2i\delta_l}-(-1)^le^{-ikr} \right] P_l(\cos\theta) \,.
\label{sc}
\end{eqnarray}
Now, this equation should be equivalent to Eq.~(\ref{basic}) with the expansion of $e^{ikz}$ in terms of partial waves given 
by Eq.~(\ref{asym1}). So, we can rewrite eqn~(\ref{basic}) as
\begin{equation}
\psi(\vec{r})_{r\rightarrow\infty}=\frac{1}{2ikr}\sum_{l=0}^{\infty}
(2l+1) \left[e^{ikr}-(-1)^le^{-ikr} \right] P_l(\cos\theta)+f_k(\theta)\frac{e^{ikr}}{r} \,.
\label{sc1}
\end{equation}
Comparing the co-efficients of ${e^{-ikr}}/{r}$ in Eqs.~(\ref{sc}) and (\ref{sc1}) one can easily get:
\begin{eqnarray}
C_l=i^l(2l+1)e^{i\delta_l} \,.
\end{eqnarray}
Once the value of $C_l$ is at hand, we can plug it into Eq.~(\ref{sc}), and then proceed to compare the co-efficients of ${e^{ikr}}/{r}$ to obtain the expression for $f_k(\theta)$. The final result is,
\begin{equation}
f_k(\theta)=\frac{1}{2ik}\sum_{l=0}^{\infty}(2l+1)(e^{2i\delta_l}-1)P_l(\cos\theta) \,.
\label{amp}
\end{equation}
Since $P_l(\cos\theta)$ serves as a complete set of basis vectors for any function of $\theta$, one can expand the scattering amplitude as follows:
\begin{equation}
f_k(\theta)=\frac{1}{k} \sum_{l=0}^{\infty}(2l+1)f_l(k)P_l(\cos\theta) \,.
\label{amp1}
\end{equation}
Comparing Eqs.~(\ref{amp}) and (\ref{amp1}) one can easily get
\begin{eqnarray}
&& f_l(k)=\frac{e^{2i\delta_l}-1}{2i} \,, \\
&\Rightarrow& f_l(k)=e^{i\delta_l} \sin\delta_l \,.
\label{flk}
\end{eqnarray}
From \Eqn{flk} it follows that
\begin{eqnarray}
|f_l(k)| \le 1 ~~~ {\rm for~ all~ values~ of~} l \,,
\label{fun}
\end{eqnarray}
or, splitting $f_l(k)$ into its real and imaginary components, one can derive the equation of the {\em unitarity circle} (see Figure~\ref{unicirc}):
\begin{eqnarray}
\Big[{\Re} f_l(k)\Big]^2 +\left[{\Im} f_l(k)-\frac{1}{2}\right]^2 = \frac{1}{4} \,.
\end{eqnarray}

Now we have learned that the expansion coefficients, $f_l(k)$, of the quantum mechanical scattering amplitude obey the unitarity conditions of \Eqn{fun}. But, till now there is very little hint that these $f_l(k)$-s are the same as the $a_l$-s of \Eqn{uni_amp}. Some intuitive arguments to make the connections will be presented shortly. Before that, to make the discussion complete, we wish to present one important result that follows from \Eqn{amp1}.
Note that, using the value of $f_l(k)$ from \Eqn{flk}, one can rewrite \Eqn{amp1} as
\begin{eqnarray}
f_k(\theta)= \frac{1}{k}  \sum_{l=0}^{\infty}(2l+1)e^{i\delta_l} \sin\delta_lP_l(\cos\theta) \,.
\label{amp2}
\end{eqnarray}
We can now find the expression for the total scattering cross-section as
\begin{eqnarray}
\sigma = \int \frac{d\sigma}{d\Omega} d\Omega = \int |f_k(\theta)|^2 d\Omega \,.
\end{eqnarray}
Using the orthonormality of the Legendre polynomials, the final result becomes
\begin{eqnarray}
\sigma = \frac{4\pi}{k^2} \sum\limits_{l=0}^{\infty} (2l+1) \sin^2\delta_l \,.
\label{sigq}
\end{eqnarray}
Looking at \Eqs{amp2}{sigq} and keeping in mind tha $P_l(1)=1$ for any $l$, one can at once realize that
\begin{eqnarray}
\sigma= \frac{4\pi}{k} \Im \left\{f_k(\theta=0)\right\} \,,
\label{optical}
\end{eqnarray}
where, $\Im \left\{f_k(\theta=0)\right\}$ is the imaginary part of the forward scattering amplitude. \Eqn{optical} is known as the {\em optical theorem} in quantum mechanics.
\subsection{Connection with Quantum Field Theory}
We shall now give a hand waving argument on how the quantum mechanical scattering amplitude is related to the Feynman amplitude in Quantum Field Theory (QFT). We know the expression for differential scattering cross-section both in quantum mechanics and in QFT. This is given by
\begin{eqnarray}
\frac{d\sigma}{d\Omega}= \underbrace{\frac{1}{64\pi^2s}|{\cm}(\theta)|^2}_{\rm QFT} = \underbrace{|f_k(\theta)|^2}_{\rm QM} \,,
\label{qmqft}
\end{eqnarray}
where, ${\cm}(\theta)$ is the Feynman amplitude for the process and $s=4E^2$ is the CM energy squared. From \Eqn{qmqft} we can make a simple-minded connection:
\begin{eqnarray}
{\cm}(\theta) =16\pi{E}f_k(\theta) \,.
\label{Mf}
\end{eqnarray}
Now, plugging the expression of $f_k(\theta)$ from \Eqn{amp1} into \Eqn{Mf} and approximating $k\approx E$ at high energies, we may write
\begin{eqnarray}
{\cm}(\theta) = 16\pi \sum_{l=0}^{\infty}(2l+1)f_l(E)P_l(\cos\theta) \,.
\label{finale}
\end{eqnarray}
Thus, comparing \Eqs{uni_amp}{finale}, one can see that $a_l$s of \Eqn{uni_amp} are the same as $f_l(k)$s of \Eqn{amp1} and both of them must obey the unitarity condition of \Eqn{fun}. Extraction of each partial wave amplitude from the Feynman amplitude will now be a straightforward task:
\begin{eqnarray}
a_l =\frac{1}{32\pi}\int_{-1}^{+1}{\cm}(\theta)P_l(\cos\theta) d(\cos\theta) \,.
\label{partial_extraction}
\end{eqnarray}
where $a_l$ are the partial wave coefficients corresponding to specific angular momentum values l. If the amplitude at tree level increases with energy then the unitary bound is violated after certain energy, then the theory becomes sick and can indicate the incompleteness of theory.
In the SM, various vector bosons scattering processes such as $W_L^\pm W_L^\mp \ra W_L^\pm  W_L^\mp$, $W_L^\pm  W_L^\pm \ra W_L^\pm  W_L^\pm$, $W_L^\pm  W_L^\mp \ra Z_L  Z_L$, $W_L^\pm  Z_L \ra W_L^\pm Z_L$ and $Z_L  Z_L \ra Z_L Z_L$ have been reviewed. It has been checked that without a Higgs boson, the unitarity condition is not fulfilled at high energies. Any extended scalar sector is in general expected to satisfy the unitarity condition, unless one can come to terms with strongly coupled physics controlling electroweak interactions at high energy. 

\section{Unitarity Constraints }
In this section, we discuss the bound from  unitarity on six-dimensional operator. We
consider all possible 2-to-2-body bosonic elastic scatterings. Every scattering amplitude can be expanded in terms of the partial waves as follows~:
\begin{equation}
\mathcal{M}(\theta) = 16 \pi\sum\limits_{\ell=0}^{\infty} a_\ell
(2\ell+1)P_\ell (\cos\theta) \,,
\end{equation}
where, $\theta$ is the scattering angle and $P_\ell (x)$ is the
Legendre polynomial of order $\ell$. The prescription is as follows:
once we calculate the Feynman amplitude of a certain $2\to 2$
scattering process, each of the partial wave amplitude ($a_\ell$), in can be extracted by using the orthonormality of the
Legendre polynomials. This technique was first developed by B. W. Lee, C. Quigg, and H. B.
Thacker for the SM ~\cite{Lee:1977eg}.In the SM they have analyzed several two body scatterings involving longitudinal gauge bosons and physical Higgs in the SM. 
The $\ell=0$ partial
wave amplitude $(a_0)$ is then extracted
from these amplitudes and cast in the form of what is called an S-matrix having different two-body eigenstates as rows and columns. The largest eigenvalue of this matrix is bounded by the unitarity constraint, $|a_0| < 1$.Using this method, we sucessfully reviewd for  SM and 2HDM case. Now, we can extend this method for 2HDM with Six-dimensional operators  2HDM. In that model, we also have the same types of two body scattering channels as in the case of 2HDM. Similary, we find out the expressions of $a_0$ for every possible $2\to 2$ scattering process and cast them in the form of an S-matrix which is constructed by taking the different two-body channels as rows and columns.\\
Firstly, we identify all the possible two-particle channels. These two-particle states are made of the fields $w_{k}^{\pm},~h_{k}$ and
$z_{k}$ corresponding to the parametrization of eqn.(1).
We consider neutral combinations out of two-particle
states (e.g., $w_{i}^{+}w_{j}^{-},~h_i h_j,~z_i z_j,~h_i z_j$) and
singly-charged two-particle states (e.g.,
$w_{i}^{+}h_j,~w_{i}^{+}z_j$). 

The neutral channel S-matrix for 2HDMEFT is a $14\times 14$ matrix with the
following two-particle states as rows and columns~:
\begin{eqnarray*}
|w_1^+w_1^->,~
|w_2^+w_2^->,~|w_1^+w_2^->,~|w_2^+w_1^->,~|\frac{h_1h_1}{\sqrt{2}}>,~|\frac{z_1z_1}{\sqrt{2}}>,
~|\frac{h_2h_2}{\sqrt{2}}>,~|\frac{z_2z_2}{\sqrt{2}}>,\end{eqnarray*}

\begin{eqnarray*}
~|h_1z_2>,~|h_2z_1>,~|z_1z_2>,~|h_1h_2>,~|h_1z_1>,~|h_2z_2>\,.
\end{eqnarray*}
The neutral sector S-matrix elements are calculated and mention in the appendix.\\
The same exercise can be repeated for the charged two-particle state
combinations. With the singly-charged state combinations, it will be a
$8\times 8$ matrix with the following two-particle states as rows and columns:
\begin{eqnarray*}
~|h_1 w_1^+>,~|h_1 w_2^+>,~|z_1 w_1^+>,
~|z_2 w_2^+>,\nn
\end{eqnarray*}
\begin{eqnarray*}
~|h_1 w_2^+>,~|h_2 w_1^+>,~|z_1 w_2^+>,
~|z_2 w_1^+>.\nn
\end{eqnarray*}
Also, for charged sector S-matrix elements are calculated and mention in the appendix.
Analytically finding the expression of eigenvalues for these matrices is too difficult. So, we solve this problem numerically and  eigenvalues corresponding to neutral and charged channels ($\Lambda_i$) will be bounded from the unitarity constraint as 
\begin{eqnarray}
|\Lambda_i| \le 8\pi \,.
\end{eqnarray}
We are  also implementing the changes in 2HDM with six-dimensional operators in 2HDMC~{\cite{Eriksson:2009ws}, so that it can be easy for the community to check unitarity in presence of dimension-six terms without calculating these effects by manually.

\subsection{$\varphi^6$ Operator}
The S-matrix for 2HDM which is block-diagonalised $22\times 22$ matrix 
and further subdivided into submatrices for  neutral channels two $6\times6$ and one $2\times2$ matrix and for charged channels two $4\times4$ matrices these feature  absent under the  $\varphi^6$ Operators. With $\varphi^6$ Operator, S-matrix will be non-diagonalised $22\times 22$ matrix. Since, the elements of S-matrix are only proportional to the $\cos\beta$, $\sin\beta$ which take value less than one and $v_{i=1,2}$ which are  much less than $f$. Thus, its  eigenvalues give insignificant unitarity bound  in comparsion to  2HDM case.

\subsection{$\varphi^4 D^2$ Operator}

We have calculated all the 2 to 2 scattering amplitudes which are enlisted in appendix A. Many of the 2 to 2 scattering amplitudes are related to each other by the virtue of Wick's theorem. \\
As discussed in previous section, the unitarity constraints demand that the eigenvalues of the scattering-matrix should be less than $8\pi$.\\
Using that condition, we directly  find the bound on $\sqrt{s}$. 
Only with $\varphi^4 D^2$ operator, the constraints on the $\sqrt{s} $ vs $f$ plane  plotted for different value of Wilson cofficients as shown in the figure 4.3.\\
\begin{figure}[h!]
 \begin{center}
\subfigure[]{
 \includegraphics[width=1.8in,height=1.8in, angle=0]{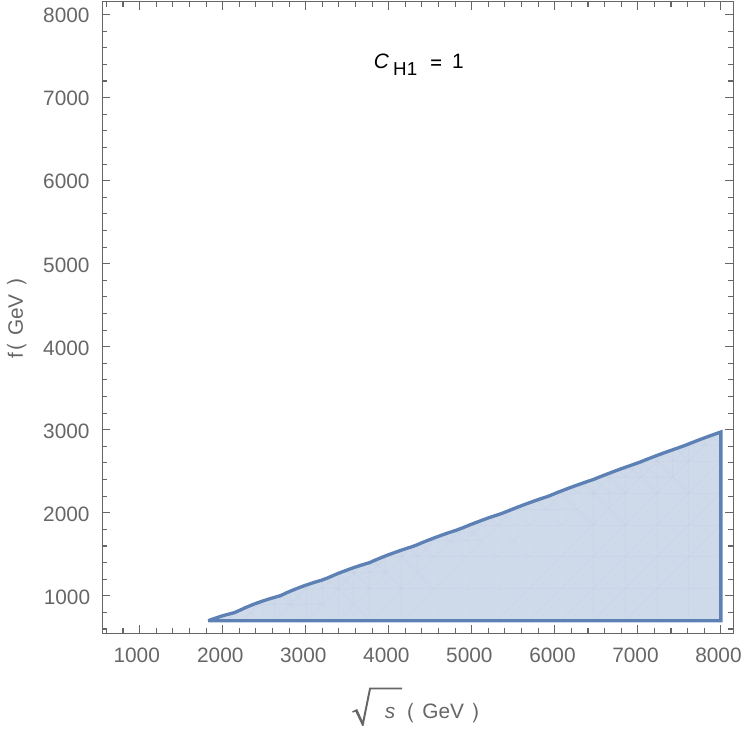}}
 \hskip 15pt
 \subfigure[]{
 \includegraphics[width=1.8in,height=1.8in, angle=0]{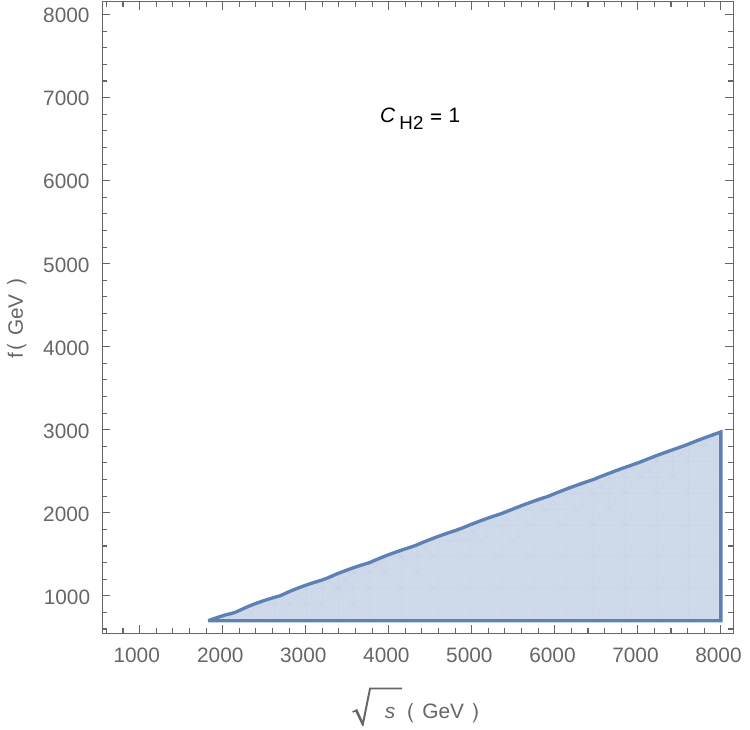}}
 \subfigure[]{
 \includegraphics[width=1.8in,height=1.8in, angle=0]{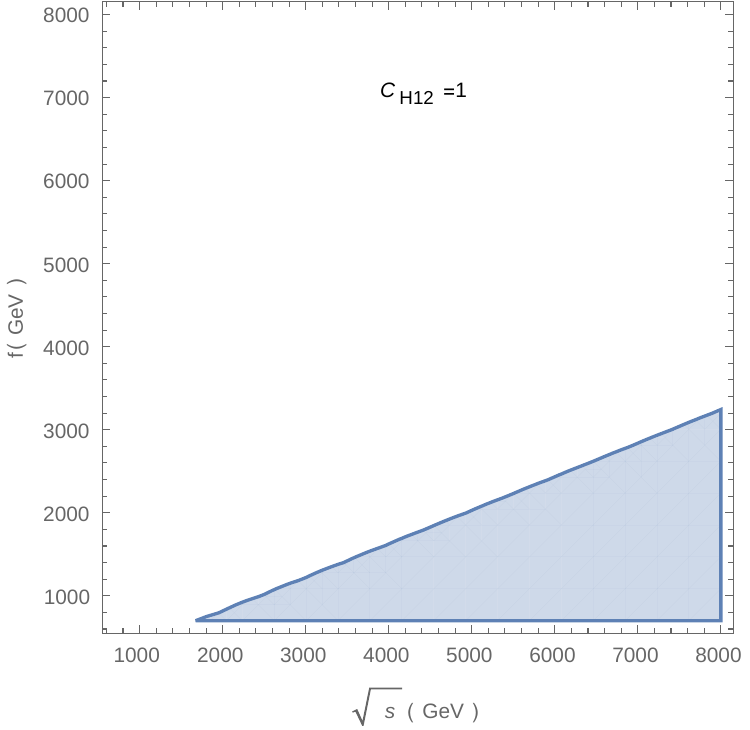}}
 \hskip 15pt
 \subfigure[]{
 \includegraphics[width=1.8in,height=1.8in, angle=0]{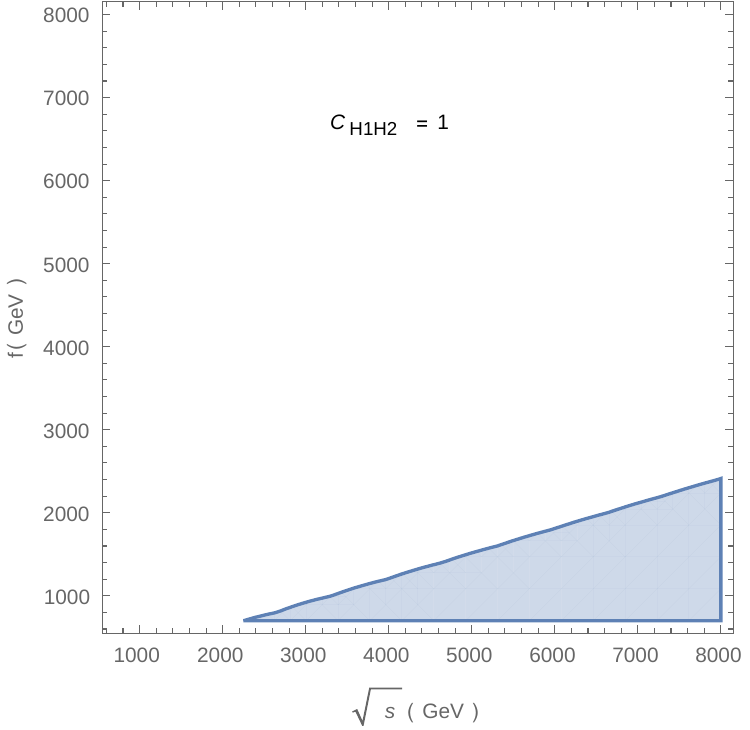}}
  \subfigure[]{
 \includegraphics[width=1.8in,height=1.8in, angle=0]{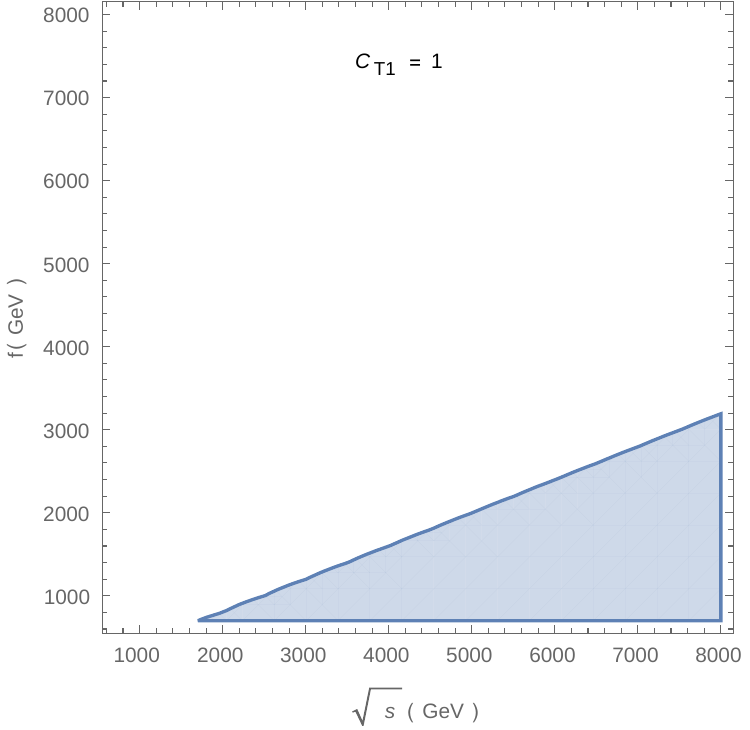}}
 \hskip 15pt
 \subfigure[]{
 \includegraphics[width=1.8in,height=1.8in, angle=0]{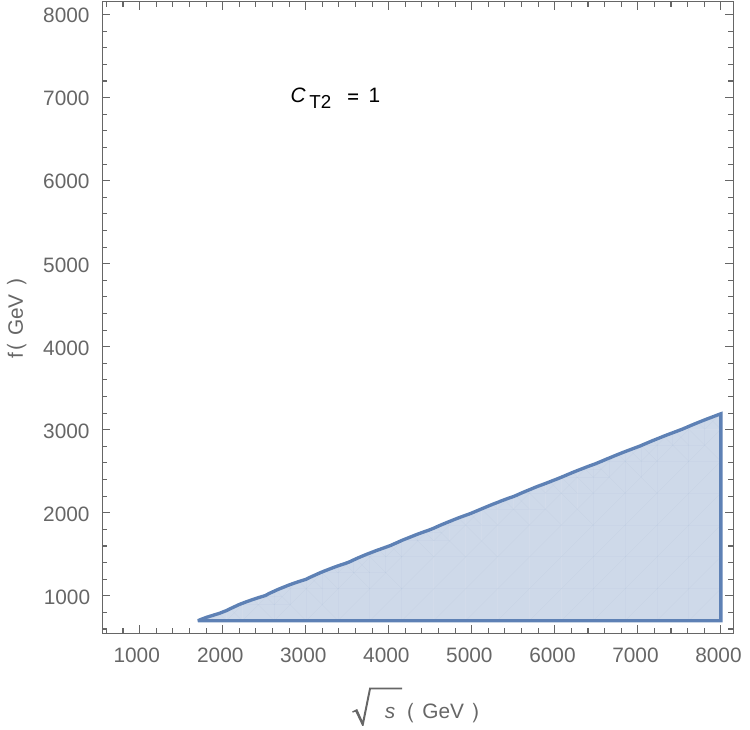}}
 \newline
 \subfigure[]{
 \includegraphics[width=1.8in,height=1.8in, angle=0]{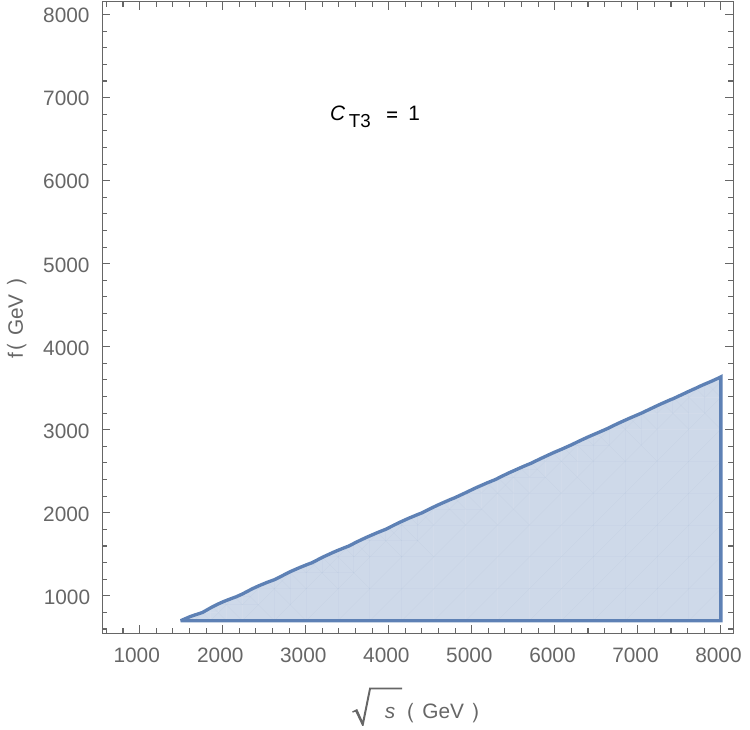}}
 \caption{The effect of $\varphi^{4} D^2$-type of operators for  2HDM. Blue regions indicate area disfavoured from the unitarity constraints for 2HDM in the presence of 6-dim operators. If we increase  $\sqrt{s} $ the lower bound on $f$ also increase and these two quantities are proportional to each other.}
\end{center} 
\label{bound}
 \end{figure}
 
\newpage


\newpage
 
From the figures we see that  under the $\varphi^4 D^2$ operator perturbative unitarity is violated around $\sqrt{s}$ =2 TeV for $f$= 1 TeV.
\subsubsection{Bounds on T-parameter violating Operator}

\begin{figure}[h!]
 \begin{center}
  \subfigure[]{
 \includegraphics[width=2.0in,height=2.0in, angle=0]{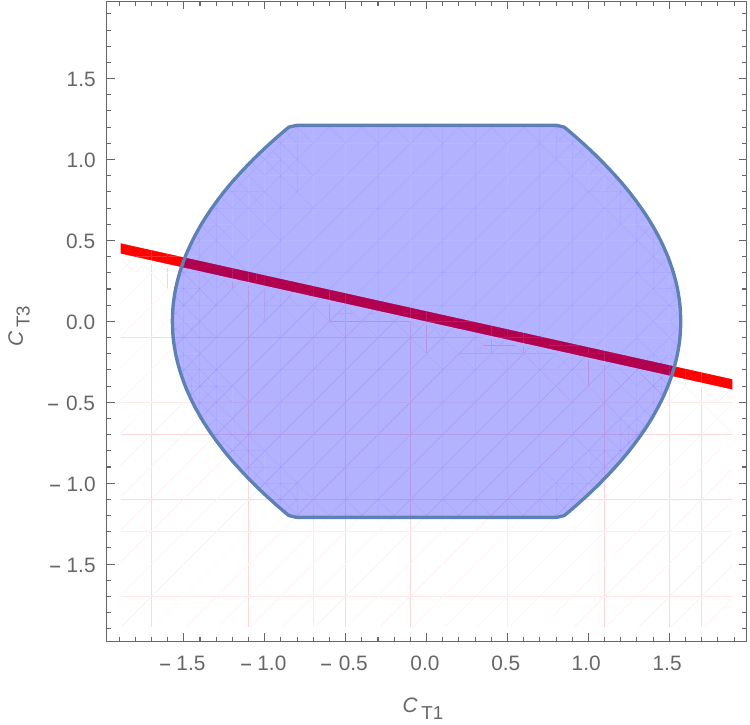}}
 \hskip 15pt
 \subfigure[]{
 \includegraphics[width=2.0in,height=2.0in, angle=0]{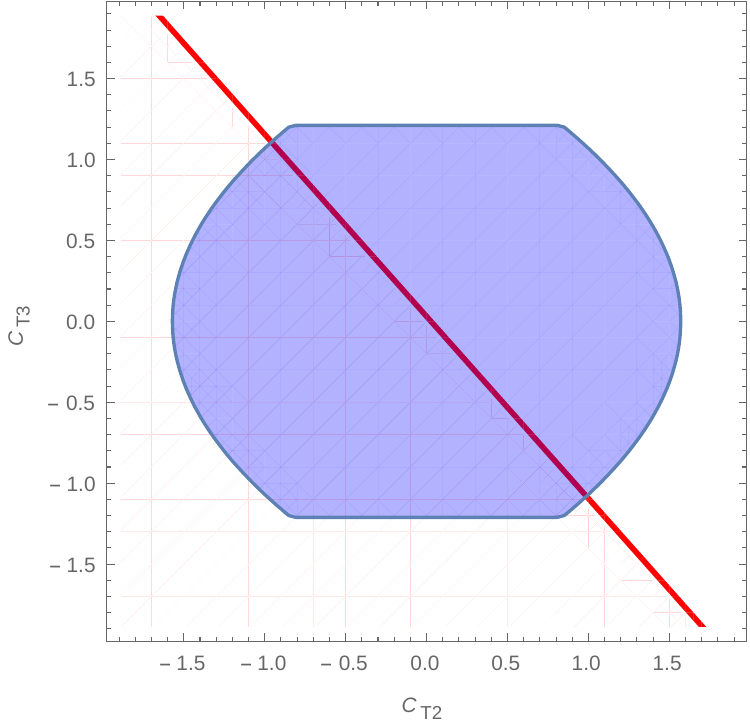}}
\caption{The unitarity limits on the Wilson coefficients, where they follow the bound $|a_0| < 1,$ (a) For $C_{T1}$ vs $C_{T3}$, (b) For $C_{T2}$ vs $C_{T3}$. Blue region  is allowed from the unitarity constraints and red region is allowed from T-parameter constraint in the Wilson coefficient plane. We have taken $\sqrt{s}$ = 2 TeV and $f$ = 1 TeV.}
 \end{center} 
 \label{bound2}
\end{figure}

From the figure 4.4, we partially constraint the combination of Wilson coefficients with help of unitarity bounds. 
\\
If all the NP fields are heavy, their effects on the 2HDM
dynamics can be parametrized by a set of higher-dimensional operators. These operators spoil the renormalisability of the effective theory and in turn can make some scattering amplitudes violate unitarity.
\\The bounds depend on the $f$ but always scale as $f^2$,  so it is easy to set a fiducial mark at $f= 1 $ TeV and show the bounds. 
\\They also depend on $\sqrt{s}$ and get stronger as  $\sqrt{s}$ increases. 
The effect of unitarity on $\varphi^{4} D^2$-type of operator  that  bound for $\sqrt{s}$ around 2 TeV, a typical parton-level energy at the LHC.
From unitarity constraints along with T-parameter measurement give stringnent bound on the Wilson coefficient.
\\

%% file: Chapter5/chapter5.tex
\chapter{Summary and Conclusions}
\label{chap:Conclusion}
\linespread{0.1}
\graphicspath{{Chapter5/}}
\pagestyle{headings}
\noindent\rule{15cm}{1.5pt} 
All the particle physics phenomena that observed in high
energy experiments at LHC  successfully described by the Standard Model. Although it cannot be the final fundamental theory of nature. Evidence of physics beyond the Standard Model appears
at cosmological scales that are not observable in the
current collider experiments. To learn about physics at higher energies, we can use theoretical arguments to study the influence of high energy physics on low energy observables.

In this thesis, we introduced a theoretical tool that can be used to study the effects of higher energy phenomena in the low energy regions that are available. From the unitarity of scatterings, which means that all possibilities add up to one, we can derive constraints on the scattering amplitudes at every order in the gauge coupling constants. The most basic scatterings are the
two-particle scatterings that can be parametrized by only two parameters, the centre-of-mass energy, and the scattering angle. After expanding the two-particle scattering amplitude in partial waves, the coefficients solely depend on the centre-of-mass energy. The perturbative unitarity limits are applied on the partial wave amplitudes which then translate into constraints on the couplings appearing in the amplitude, or on the energy. When an amplitude grows with the energy, then it will violate unitarity at some point. This is the cut-off of the theory, where it loses predictive power. If we want to create a theory that is valid up to arbitrarily high energies, it must respect perturbative unitarity in all scattering amplitude at all energies.\\

Studying the scattering processes using perturbative unitarity constraints, we can draw limits on the parameters of the theory or establish the validity scale of the theory where new physics should appear to correct the unitarity violating processes. Also, depending on which scatterings violate unitarity, we can forecast the new physics that unitarizes the process. That makes perturbative unitarity a useful tool for studying new theories. \\
In this thesis,  extended scalar sectors of the SM have been discussed. The SM has been extended with a SU(2) doublet. It has been considered that the extra scalar fields transform under the same standard model gauge group.
In that model, it has been considered that both the SM doublet and the extra scalar field are responsible for the electroweak symmetry breaking, i.e., both the neutral  CP-even component of SM doublet and additional scalar fields are getting vacuum expectation values. The different extended scalar sectors include various kinds of scalar particles such as charged, or neutral CP-even and CP-odd scalar(s). These scalars can couple to the vector bosons.
In this point of view, various kinds of vector boson scattering processes have been reviewed for 2HDM.\\
The 2HDMEFT study was motivated by SMEFT. From that, 
 we introduced an effective model that extends 2HDM with six-dimensional bosonic operators. We calculated different two-to-two scatterings which were growing with the centre-of-mass energy. After applying the perturbative unitarity constraints, we obtained that the model has high validity scale. In the end, we find out constraints from unitarity along with T-parameter measurement give stringent bound on the Wilson coefficient. We partially constraint the combination of Wilson coefficients with the help of unitarity bounds. \\
What we can see from the  studies is that a perturbative unitarity is a useful tool in complementing the experimental constraints on the parameter space and establish the limits of effective models.

%% file: AppendixA/appendixA.tex
\graphicspath{{Appendix}}
\begin{appendices}
\chapter{ Analytical expressions of S-matrix  \label{sec:wwscatdetails}}
\pagestyle{headings}
\pagenumbering{arabic}

\textbf{The Potential}
\newline
The  total potential is looks like: 
$ V(\varphi_1,\varphi_2) + \mathcal{L}_{\varphi^6}$. Here, $V(\varphi_1,\phi_2)$ is given in eqn.(2.2.4)  and
\begin{eqnarray*}
\mathcal{L}_{\varphi^6} &=& \frac{1}{f^2}\Big[c_{111} |\varphi_1|^6 + c_{222} |\varphi_2|^6 
+ c_{112} |\varphi_1|^4 |\varphi_2|^2 + c_{122} |\varphi_1|^2 |\varphi_2|^4 \\
&&+ c_{(1221)1} |\varphi_1^{\dagger} \varphi_2|^2 |\varphi_1|^2 + c_{(1221)2} |\varphi_1^{\dagger} \varphi_2|^2 |\varphi_2|^2 \\
&&+ c_{(1212)1} ((\varphi_1^{\dagger} \varphi_2)^2 + h.c.) |\varphi_1|^2 
+ c_{(1212)2}  ((\varphi_1^{\dagger} \varphi_2)^2 + h.c.) |\varphi_2|^2 \\
&&+ \textcolor{blue}{c_{(1221)12} |\varphi_1^{\dagger} \varphi_2|^2 (\varphi_1^{\dagger} \varphi_2 + h.c.)} 
+ \textcolor{blue}{c_{11(12)} |\varphi_1|^4 (\varphi_1^{\dagger} \varphi_2 + h.c.)} \\
&&+ \textcolor{blue}{c_{22(12)} |\varphi_2|^4 (\varphi_1^{\dagger} \varphi_2 + h.c.)} 
+ \textcolor{blue}{c_{12(12)} |\varphi_1|^2 |\varphi_2|^2 (\varphi_1^{\dagger} \varphi_2 + h.c.)}\\ 
&&+ \textcolor{blue}{c_{121212} (\varphi_1^{\dagger} \varphi_2 + h.c.)^3}  \Big].
\end{eqnarray*}
Here, we have marked the $Z_2$-violating operators in blue colour. The minimisation conditions of this potential are: 
\begin{eqnarray*}
&&\frac{3}{4}v_1^4 c_{111} + \frac{v_1^2 v_2^2}{2}c_{112} + \frac{v_2^4}{4}c_{122}+ v_1^2 v_2^2 c_{(1212)1} + \frac{v_2^4}{2}c_{(1212)2}+ \frac{v_1^2 v_2^2}{2}c_{(1221)1}  + \frac{v_2^4}{4}c_{(1221)2} \nn\\
&&+ \frac{3}{4}v_1 v_2^3 c_{(1221)12} + \frac{5}{4}v_1^3 v_2 c_{11(12)} + \frac{3}{4}v_1 v_2^3 c_{12(12)} + \frac{v_2^5}{4 v_1}c_{22(12)} + 3 v_1 v_2^3 c_{121212}  = 0 \, ,\\
\text{and}\\
&&\frac{3}{4}v_{2}^4 c_{222} + \frac{v_1^4}{4}c_{112} + \frac{v_1^2 v_2^2}{2}c_{122} +\frac{v_1^4}{2}c_{(1212)1} + v_1^2 v_2^2 c_{(1212)2}+ \frac{v_1^4}{4} c_{(1221)1} + \frac{v_1^2 v_2^2}{2}c_{(1221)2} \nn\\
&&+\frac{3}{4}v_1^3 v_2 c_{(1221)12} + \frac{5}{4} v_1 v_2^3 c_{22(12)} + \frac{3}{4} v_1^3 v_2 c_{(12)12} + 3 v_1^3 v_2 c_{121212} = 0\, .
\end{eqnarray*}
\newpage
\textbf{$\varphi^4 D^2$ Operators}
\newline
These operators lead to the rescaling of the kinetic terms of all the Higgs fields, without
the charged scalars. Such effects should be taken care of by appropriate field redefinitions,
which lead to the scaling of the couplings of the SM-like Higgs.
\begin{eqnarray*}
\mathcal{L}_{\varphi^4 D^2} &=& \frac{1}{f^2}\big[C_{H1} O_{H1} + C_{H2} O_{H2}
+ C_{H12} O_{H12} + C_{H1H2} O_{H1H2}\\
&& +\textcolor{blue} {C_{H1H12}} \textcolor{blue}{O_{H1H12}} + \textcolor{blue}{C_{H2H12}} \textcolor{blue}{O_{H2H12}}+  C_{T1} O_{T1}+ C_{T2} O_{T2}+ C_{T3} O_{T3}\\
&&+\textcolor{blue} {C_{T4}} \textcolor{blue}{O_{T4}}+ \textcolor{blue}{C_{T5}} \textcolor{blue}{O_{T5}}\big],
\end{eqnarray*}
where,
\begin{eqnarray*}
\label{ops}
O_{H1} &=& (\partial_{\m}|\varphi_1|^2)^2, \hspace{10pt} O_{H2} = (\partial_{\m}|\varphi_2|^2)^2,\hspace{10pt} O_{H12} = (\partial_{\m}(\varphi_1^{\dagger} \varphi_2 + h.c.))^2, \\
O_{H1H2} &=& \partial_{\m}|\varphi_1|^2 \partial^{\m}|\varphi_2|^2, O_{H1H12} = \partial_{\m}|\varphi_1|^2\partial^{\m}(\varphi_1^{\dagger} \varphi_2 + h.c.),
\\ && O_{H2H12} = \partial_{\m}|\varphi_2|^2\partial^{\m}(\varphi_1^{\dagger} \varphi_2 + h.c.).
\end{eqnarray*}
Operators $O_{H1H12}$ and $O_{H2H12}$ are odd under the $Z_2$-symmetry, whereas the rest are even.

Here, we neglect the contribution from  $\varphi^2 X^2$ and $\varphi^2 D^2 X^2$ types operator becuase Constraints from electroweak precision test(EWPT) for these operators insignificant  for our purpose.
After including, the higher dimensional  operators to the 2HDM tree-level S-matrix elements looks like:
\begin{eqnarray*}
\mathcal{M}&=&\mathcal{M}_{tree-level}+\mathcal{M}_{6-dim},\nn
\end{eqnarray*}
In this case, the neutral channel S-matrix 
$14\times14$ with the following two particle states as rows and columns:
\begin{eqnarray*}
1\equiv|w_1^+w_1^->,
~2\equiv|w_2^+w_2^->,
~3\equiv|\frac{z_1z_1}{\sqrt{2}}>,
~4\equiv|\frac{z_2z_2}{\sqrt{2}}>,
~5\equiv|\frac{h_1h_1}{\sqrt{2}}>,
\nn
\end{eqnarray*}
\begin{eqnarray*}
~6\equiv|\frac{h_2h_2}{\sqrt{2}}>,~7\equiv|w_1^+w_2^->,
~8\equiv|w_2^+w_1^->,
~9\equiv|h_1z_2>,
~10\equiv|h_2z_1>,\nn
\end{eqnarray*}
\begin{eqnarray*}
~11\equiv|z_1z_2>,
~12\equiv|h_1h_2>,
~13\equiv|h_1z_1>,
~14\equiv|h_2z_2>.
\end{eqnarray*}

The elements of neutral sector $14\times 14$ matrix are given by
\begin{eqnarray*}
\mathcal{M}^{N}_{1,1}&=&4 \left(\frac{3 {c_{111}} {v_1}^2}{2 f^2}+\frac{{c_{112}} {v_2}^2}{2 f^2}+\lambda _1+\lambda _3\right)+\frac{2 u {C_{H1}} }{f^2},\nn\\
\mathcal{M}^{N}_{1,2}&=&\frac{{c_{112}} {v_1}^2}{f^2}+\frac{{c_{122}} {v_2}^2}{f^2}+\frac{{c_{(1221)1}} {v_1}^2}{2 f^2}+\frac{{c_{(1212)2}} {v_2}^2}{2 f^2}+\frac{2 s {C_{H12}} }{f^2}+\frac{ s {C_{H1H2}} }{2 f^2}\\
 && +2 \lambda _3+\frac{\lambda _5}{2}+\frac{\lambda _6}{2},\nn\\
\mathcal{M}^{N}_{1,3}&=& \sqrt{2} \left(\frac{3 {c_{111}} {v_1}^2}{2 f^2}+\frac{{c_{112}} {v_2}^2}{2 f^2}+\frac{{c_{(1221)1}} {v_2}^2}{4 f^2}-\frac{{c_{1212)1}} {v_2}^2}{2 f^2}+\lambda _1+\lambda _3\right)+\frac{2 s{C_{H1}} }{f^2},\nn\\
\mathcal{M}^{N}_{1,4}&=&
\sqrt{2} \left(\frac{9 {c_{111}} {v_1}^2}{2 f^2}+\frac{{c_{112}} {v_2}^2}{2 f^2}+\frac{{c_{(1221)1}} {v_2}^2}{4 f^2}+\frac{{c_{(1212)1}}
{v_2}^2}{2 f^2}+\lambda _1+\lambda _3\right)+\frac{2 s C_{H1}} {f^2},\nn\\
\mathcal{M}^{N}_{1,5}&=&
\sqrt{2} \left(\frac{c_{112} {v_1}^2}{2 f^2}+\frac{c_{122} {v_2}^2}{2 f^2}+\frac{c_{(1221)1} {v_1}^2}{4 f^2}-\frac{c_{(1212)1} {v_1}^2}{2 f^2}+\lambda _3+\frac{\lambda _4}{2}\right)+\frac{ s C_{H1H2} }{f^2},\nn\\
\mathcal{M}^{N}_{1,6}&=&
\sqrt{2} \left(\frac{c_{112} {v_1}^2}{2 f^2}+\frac{3 c_{122} {v_2}^2}{2 f^2}+\frac{c_{(1221)1} {v_1}^2}{4 f^2}+\frac{c_{(1212)1} {v_1}^2}{2 f^2}+\lambda _3+\frac{\lambda _4}{2}\right)+\frac{ s C_{H1H2} }{2 f^2},\nn\
\mathcal{M}^{N}_{1,7}&=&
\frac{c_{(1221)1} v_1 v_2}{2 f^2}+\frac{c_{(1212)1} v_1 v_2}{f^2},\nn\
\mathcal{M}^{N}_{1,8}&=&
\frac{c_{(1221)1} v_1 v_2}{2 f^2}+\frac{c_{(1212)1} v_1 v_2}{f^2},\nn\
\mathcal{M}^{N}_{1,9}&=&\mathcal{M}^{N}_{1,10}=0,\nn\\
\mathcal{M}^{N}_{1,11}&=&\frac{2 c_{(1212)1} v_1 v_2}{f^2},\nn\\
\mathcal{M}_{1,12}&=&\frac{2 c_{112} v_1 v_2}{f^2}+\frac{c_{(1221)1} v_1 v_2}{f^2}+\frac{2 c_{(1212)1} v_1 v_2}{f^2},\nn\\
\mathcal{M}^{N}_{1,13}&=&\mathcal{M}^{N}_{1,14}=0,\nn\\
\mathcal{M}^{N}_{2,2}&=&  \frac{2 C_{H_2} u}{f^2}+ 4 \left(\frac{c_{122} v_1^2}{2 f^2}+\frac{3 c_{222} v_2^2}{2 f^2}+\lambda _2+\lambda _3\right),\nn\\
\mathcal{M}^{N}_{2,3}&=& \frac{2 C_{H1H2} s}{f^2}+\sqrt{2} \left(\frac{c_{112} v_1^2}{2 f^2}+\frac{c_{122} v_2^2}{2 f^2}+\frac{c_{(1221)2} v_2^2}{4 f^2}-\frac{c_{(1212)2} v_2^2}{2 f^2}+\lambda _3+\frac{\lambda _4}{2}\right)\nn\\
\end{eqnarray*}
\begin{eqnarray*}
\mathcal{M}^{N}_{2,4}&=& \frac{ C_{H_1H_2} s}{2 f^2}+\sqrt{2} \left(\frac{3 c_{112} v_1^2}{2 f^2}+\frac{c_{122} v_2^2}{2 f^2}+\frac{c_{(1221)2} v_2^2}{4 f^2}+\frac{c_{(1212)2} v_2^2}{2 f^2}+\lambda _3+\frac{\lambda _4}{2}\right)\nn\\
\mathcal{M}^{N}_{2,5}&=&\frac{2 C_{H_2} s}{f^2}+\sqrt{2} \left(\frac{c_{122} v_1^2}{2 f^2}+\frac{c_{(1221)2} v_1^2}{4 f^2}-\frac{c_{(1212)2} v_1^2}{2 f^2}+\frac{3 c_{222} v_2^2}{2 f^2}+\lambda _2+\lambda _3\right)\nn
\mathcal{M}^{N}_{2,6}&=&\frac{2 C_{H_2} s}{f^2}+\sqrt{2} \left(\frac{c_{122} v_1^2}{2 f^2}+\frac{c_{(1221)2} v_1^2}{4 f^2}+\frac{c_{(1212)2} v_1^2}{2 f^2}+\frac{9 c_{222} v_2^2}{2 f^2}+\lambda _2+\lambda _3\right)\nn\\\mathcal{M}^{N}_{2,7}&=& \left(\frac{c_{(1221)2} v_1 v_2}{2 f^2}+\frac{c_{(1212)2} v_1 v_2}{f^2}\right),\nn\\
\mathcal{M}^{N}_{2,8}&=& \left(\frac{c_{(1221)2} v_1 v_2}{2 f^2}+\frac{c_{(1212)2} v_1 v_2}{f^2}\right)\nn\\
\mathcal{M}^{N}_{2,9}&=&\mathcal{M}^{N}_{2,10}=0,\nn\\
\mathcal{M}^{N}_{2,11}&=&\frac{2 c_{(1212)2} v_1 v_2}{f^2},\nn\\
\mathcal{M}^{N}_{2,12}&=&\frac{2 c_{122} v_1 v_2}{f^2}+\frac{c_{(1221)2} v_1 v_2}{f^2}+\frac{2 c_{(1212)2} v_1 v_2}{f^2},\nn\\
\mathcal{M}^{N}_{2,13}&=&\mathcal{M}^{N}_{2,14}=0,\nn\\
\mathcal{M}^{N}_{3,3}&=& 12 \left(\frac{3 c_{111} v_1^2}{8 f^2}+\frac{c_{112} v_2^2}{8 f^2}+\frac{c_{(1221)1} v_2^2}{8 f^2}-\frac{c_{(1212)1} v_2^2}{4 f^2}+\frac{\lambda _1}{4}+\frac{\lambda _3}{4}\right),\nn\\
\mathcal{M}^{N}_{3,4}&=&\frac{2 C_{H_1} s}{f^2}+\frac{36 c_{111} v_1^2+4 c_{112} v_2^2+4 c_{(1221)1} v_2^2+8 f^2 \lambda _1+8 f^2 \lambda _3}{8 f^2},\nn\\
\mathcal{M}^{N}_{3,5}&=& \frac{C_{H_1H_2} s}{f^2}+\frac{2 c_{112} v_1^2+2 c_{(1221)1} v_1^2+2 c_{122} v_2^2+2 c_{(1221)2} v_2^2+4 f^2 \lambda _3+2 f^2 \lambda _5}{4 f^2},\nn\\
\mathcal{M}^{N}_{3,6}&=& \frac{C_{H_1H_2} s}{f^2}+\frac{2 c_{112} v_1^2+2 c_{(1221)1} v_1^2+6 c_{122} v_2^2+6 c_{(1221)2} v_2^2-12 c_{(1212)2} v_2^2+4 f^2 \lambda _3+2 f^2 \lambda _6}{4 f^2},\nn\\
\mathcal{M}^{N}_{3,7}&=&\mathcal{M}^{N}_{3,8}=\mathcal{M}^{N}_{3,9}=\mathcal{M}^{N}_{3,10}=\mathcal{M}^{N}_{3,11}=\mathcal{M}^{N}_{3,12}=\mathcal{M}^{N}_{3,13}=\mathcal{M}^{N}_{3,14}=0,\nn\\
\end{eqnarray*}

\begin{eqnarray*}
\mathcal{M}^{N}_{4,4}&=& 12 \left(\frac{c_{112} v_2^2}{8 f^2}+\frac{c_{(1221)1} v_2^2}{8 f^2}+\frac{c_{(1212)1} v_2^2}{4 f^2}+\frac{15 \text{C$\_$} v_1^2}{8 f^2}+\frac{\lambda _1}{4}+\frac{\lambda _3}{4}\right),\nn\\
\mathcal{M}^{N}_{4,5}&=&\frac{C_{H_1H_2} s}{f^2}+\frac{4 c_{112} v_2^2+4 c_{(1221)1} v_2^2+36 \text{C$\_$} v_1^2+8 f^2 \lambda _1+8 f^2 \lambda _3}{8 f^2},\nn\\
\mathcal{M}^{N}_{4,6}&=&\frac{C_{H_1H_2} s}{f^2}+\frac{2 c_{112} v_1^2+2 c_{(1221)1} v_1^2+2 c_{122} v_2^2+2 c_{(1212)1} v_2^2+4 f^2 \lambda _3+2 f^2 \lambda _5}{4 f^2},\nn\\
\mathcal{M}^{N}_{4,7}&=&\mathcal{M}^{N}_{4,8}=\mathcal{M}^{N}_{4,9}=\mathcal{M}^{N}_{4,10}=\mathcal{M}^{N}_{4,11}=\mathcal{M}^{N}_{4,12}=\mathcal{M}^{N}_{4,13}=\mathcal{M}^{N}_{4,14}=0,\nn
\mathcal{M}^{N}_{5,5}&=&12 \left(\frac{c_{122} v_1^2}{8 f^2}+\frac{c_{(1212)1} v_1^2}{8 f^2}-\frac{c_{(1212)2} v_1^2}{4 f^2}+\frac{3 c_{222} v_2^2}{8 f^2}+\frac{\lambda _2}{4}+\frac{\lambda _3}{4}\right),\nn\\
\mathcal{M}^{N}_{5,6}&=&\frac{2 C_{H_2} s}{f^2}+\frac{4 c_{122} v_1^2+4 c_{(1212)1} v_1^2+36 c_{222} v_2^2+8 f^2 \lambda _2+8 f^2 \lambda _3}{8 f^2},\nn\\
\mathcal{M}^{N}_{5,7}&=&\mathcal{M}^{N}_{5,8}=\mathcal{M}^{N}_{5,9}=\mathcal{M}^{N}_{5,10}=\mathcal{M}^{N}_{5,11}=\mathcal{M}^{N}_{5,12}=\mathcal{M}^{N}_{5,13}=\mathcal{M}^{N}_{5,14}=0,\nn\\
\mathcal{M}^{N}_{6,6}&=&12 \left(\frac{c_{122} v_1^2}{8 f^2}+\frac{c_{(1212)1} v_1^2}{8 f^2}+\frac{c_{(1212)2} v_1^2}{4 f^2}+\frac{15 c_{222} v_2^2}{8 f^2}+\frac{\lambda _2}{4}+\frac{\lambda _3}{4}\right),\nn\\
\mathcal{M}^{N}_{6,7}&=&\mathcal{M}^{N}_{6,8}=\mathcal{M}^{N}_{6,9}=\mathcal{M}^{N}_{6,10}=\mathcal{M}^{N}_{6,11}=\mathcal{M}^{N}_{6,12}=\mathcal{M}^{N}_{6,13}=\mathcal{M}^{N}_{6,14}=0,\nn
\mathcal{M}^{N}_{7,7}&=& \frac{C_{H_1H_2} t}{f^2}+4 \left(\frac{c_{112} v_1^2}{f^2}+\frac{c_{(1221)1} v_1^2}{2 f^2}+\frac{c_{122} v_2^2}{f^2}+\frac{c_{(1212)1} v_2^2}{2 f^2}+2 \lambda _3+\frac{\lambda _5}{2}+\frac{\lambda _6}{2}\right),\nn\\
\mathcal{M}^{N}_{7,8}&=&\frac{C_{H_{12}} t}{f^2}+\frac{c_{(1212)1} v_1^2}{2 f^2}+\frac{c_{(1212)2} v_2^2}{2 f^2}+\frac{\lambda _5}{4}-\frac{\lambda _6}{4},\nn
\mathcal{M}^{N}_{7,9}&=& \frac{3 i c_{(1221)1} v_1^2}{4 f^2}-\frac{3 i c_{(1212)1} v_1^2}{2 f^2}+\frac{i c_{(1212)1} v_2^2}{4 f^2}-\frac{i c_{(1212)2} v_2^2}{2 f^2}+\frac{i \lambda _6}{2}-\frac{i \lambda _4}{2},\nn\\
\mathcal{M}^{N}_{7,10}&=&\frac{i c_{(1212)1} v_1^2}{2 f^2}-\frac{i c_{(1221)1} v_1^2}{4 f^2}+\frac{3 i c_{(1212)2} v_2^2}{2 f^2}-\frac{3 i c_{(1212)1} v_2^2}{4 f^2}+\frac{i \lambda _4}{2}-\frac{i \lambda _6}{2},\nn\\
\mathcal{M}^{N}_{7,11}&=&\frac{2 C_{H_{12}} s}{f^2}+\frac{c_{(1221)1} v_1^2}{4 f^2}+\frac{c_{(1212)1} v_1^2}{2 f^2}+\frac{c_{(1212)1} v_2^2}{4 f^2}+\frac{c_{(1212)2} v_2^2}{2 f^2}+\frac{\lambda _5}{2}-\frac{\lambda _4}{2},\nn\\
\mathcal{M}^{N}_{7,12}&=&\frac{2 C_{H_{12}} s}{f^2}+\frac{3 c_{(1221)1} v_1^2}{4 f^2}+\frac{3 c_{(1212)1} v_1^2}{2 f^2}+\frac{3 c_{(1212)1} v_2^2}{4 f^2}+\frac{3 c_{(1212)2} v_2^2}{2 f^2}+\frac{\lambda _5}{2}-\frac{\lambda _4}{2},\nn\\
\mathcal{M}^{N}_{7,13}&=&\frac{i c_{(1221)1} v_1 v_2}{2 f^2}-\frac{i c_{(1212)1} v_1 v_2}{f^2},\nn\\
\end{eqnarray*}
\begin{eqnarray*}
\mathcal{M}^{N}_{7,14}&=&\frac{i c_{(1212)2} v_1 v_2}{f^2}-\frac{i c_{(1212)1} v_1 v_2}{2 f^2},\nn
\mathcal{M}^{N}_{8,8}&=&\frac{C_{H1H2} t}{f^2}+4 \left(\frac{c_{112} v_1^2}{f^2}+\frac{c_{(1221)1} v_1^2}{2 f^2}+\frac{c_{122} v_2^2}{f^2}+\frac{c_{(1212)1} v_2^2}{2 f^2}+2 \lambda _3+\frac{\lambda _5}{2}+\frac{\lambda _6}{2}\right),\nn\\
\mathcal{M}^{N}_{8,9}&=&\frac{3 i c_{(1212)1} v_1^2}{2 f^2}-\frac{3 i c_{(1221)1} v_1^2}{4 f^2}+\frac{i c_{(1212)2} v_2^2}{2 f^2}-\frac{i c_{(1212)1} v_2^2}{4 f^2}+\frac{i \lambda _4}{2}-\frac{i \lambda _6}{2},\nn\\
\mathcal{M}^{N}_{8,10}&=&\frac{i c_{(1221)1} v_1^2}{4 f^2}-\frac{i c_{(1212)1} v_1^2}{2 f^2}+\frac{3 i c_{(1212)1} v_2^2}{4 f^2}-\frac{3 i c_{(1212)2} v_2^2}{2 f^2}+\frac{i \lambda _6}{2}-\frac{i \lambda _4}{2},\nn\\
\mathcal{M}^{N}_{8,11}&=&\frac{2 C_{H12} s}{f^2}+\frac{i c_{(1221)1} v_1^2}{4 f^2}-\frac{i c_{(1212)1} v_1^2}{2 f^2}+\frac{3 i c_{(1212)1} v_2^2}{4 f^2}-\frac{3 i c_{(1212)2} v_2^2}{2 f^2}+\frac{i \lambda _6}{2}-\frac{i \lambda _4}{2},\nn\\
\mathcal{M}^{N}_{8,12}&=&\frac{2 C_{H12} s}{f^2}+ \frac{c_{(1221)1} v_1^2}{4 f^2}+\frac{c_{(1212)1} v_1^2}{2 f^2}+\frac{c_{(1212)1} v_2^2}{4 f^2}+\frac{c_{(1212)2} v_2^2}{2 f^2}+\frac{\lambda _5}{2}-\frac{\lambda _4}{2},\nn\\
\mathcal{M}^{N}_{8,13}&=&\frac{i c_{(1212)1} v_1 v_2}{f^2}-\frac{i c_{(1221)1} v_1 v_2}{2 f^2},\nn\\
\mathcal{M}^{N}_{8,14}&=&\frac{i c_{(1212)1} v_1 v_2}{2 f^2}-\frac{i c_{(1212)2} v_1 v_2}{f^2},\nn\\
\mathcal{M}^{N}_{9,9}&=& \frac{C_{H1H2} t}{f^2} +4 \left(\frac{3 c_{112} v_1^2}{4 f^2}+\frac{3 c_{(1221)1} v_1^2}{4 f^2}-\frac{3 c_{(1212)1} v_1^2}{2 f^2}+\frac{c_{122} v_2^2}{4 f^2}+\frac{c_{(1212)1} v_2^2}{4 f^2}+\frac{\lambda _3}{2}+\frac{\lambda _6}{4}\right),\nn\\
\mathcal{M}^{N}_{9,10}&=&\frac{2 C_{H12} t}{f^2}+\frac{3 c_{(1212)1} v_1^2}{f^2}+\frac{3 c_{(1212)2} v_2^2}{f^2}+\frac{\lambda _5}{2}-\frac{\lambda _6}{2},\nn\\
\mathcal{M}^{N}_{9,11}&=&\mathcal{M}^{N}_{9,12}=0,\nn\\
\mathcal{M}^{N}_{9,13}&=&
\frac{6 c_{(1212)1} v_1 v_2}{f^2},\nn\\
\mathcal{M}_{9,14}&=&
2 \left(\frac{c_{122} v_1 v_2}{f^2}+\frac{c_{(1212)1} v_1 v_2}{f^2}\right),\nn\\
\end{eqnarray*}
\small{
\begin{eqnarray*}
\mathcal{M}^{N}_{10,10}&=&\frac{2 C_{H12} t}{f^2}+4 \left(\frac{c_{112} v_1^2}{4 f^2}+\frac{c_{(1221)1} v_1^2}{4 f^2}+\frac{3 c_{122} v_2^2}{4 f^2}+\frac{3 c_{(1212)1} v_2^2}{4 f^2}-\frac{3 c_{(1212)2} v_2^2}{2 f^2}+\frac{\lambda _3}{2}+\frac{\lambda _6}{4}\right),\nn\\
\mathcal{M}^{N}_{10,11}&=&\mathcal{M}^{N}_{10,12}=0,\nn\\
\mathcal{M}^{N}_{10,13}&=&2 \left(\frac{c_{112} v_1 v_2}{f^2}+\frac{c_{(1221)1} v_1 v_2}{f^2}\right),\nn
\mathcal{M}^{N}_{10,14}&=&\frac{3 c_{(1212)2} v_1 v_2}{f^2},\nn\\
\mathcal{M}^{N}_{11,11}&=&\frac{C_{H12} t}{f^2}+\frac{C_{H1H2} t}{f^2}+4 \left(\frac{c_{112} v_1^2}{4 f^2}+\frac{c_{(1221)1} v_1^2}{4 f^2}+\frac{c_{122} v_2^2}{4 f^2}+\frac{c_{(1212)1} v_2^2}{4 f^2}+\frac{\lambda _3}{2}+\frac{\lambda _5}{4}\right),\nn\\
\mathcal{M}^{N}_{11,12}&=&\frac{2 C_{H12} s}{f^2}+\frac{3 c_{(1212)1} v_1^2}{f^2}+\frac{3 c_{(1212)2} v_2^2}{f^2}+\frac{\lambda _5}{2}-\frac{\lambda _6}{2},\nn\\
\mathcal{M}^{N}_{11,13}&=&
\mathcal{M}^{N}_{11,14}=0,\nn\\
\mathcal{M}^{N}_{12,12}&=&\frac{C_{H12} t}{f^2}+\frac{C_{H1H2} t}{f^2}+4 \left(\frac{3 c_{112} v_1^2}{4 f^2}+\frac{3 c_{(1221)1} v_1^2}{4 f^2}+\frac{3 c_{(1212)1} v_1^2}{2 f^2}\right)\nn\\
&&+4 \left(\frac{3 c_{122} v_2^2}{4 f^2}+\frac{3 c_{(1212)1} v_2^2}{4 f^2}+\frac{3 c_{(1212)2} v_2^2}{2 f^2}+\frac{\lambda _3}{2}+\frac{\lambda _5}{4}\right),\nn\\
\mathcal{M}^{N}_{12,13}&=&
\mathcal{M}^{N}_{12,14}=0,\nn\\
\mathcal{M}^{N}_{13,13}&=&\frac{2 C_{H1} t}{f^2}+4 \left(\frac{c_{112} v_2^2}{4 f^2}+\frac{c_{(1221)1} v_2^2}{4 f^2}+\frac{9  c_{111} v_1^2}{4 f^2}+\frac{\lambda _1}{2}+\frac{\lambda _3}{2}\right),\nn\\
\mathcal{M}^{N}_{13,14}&=&\frac{2 C_{H12} t}{f^2}+\frac{3 c_{(1212)1} v_1^2}{f^2}+\frac{3 c_{(1212)2} v_2^2}{f^2}+\frac{\lambda _5}{2}-\frac{\lambda _6}{2},\nn\\
\mathcal{M}^{N}_{14,14}&=&\frac{2 C_{H2} t}{f^2}+4 \left(\frac{c_{122} v_1^2}{4 f^2}+\frac{c_{(1212)1} v_1^2}{4 f^2}+\frac{9 c_{222} v_2^2}{4 f^2}+\frac{\lambda _2}{2}+\frac{\lambda _3}{2}\right).\nn
\end{eqnarray*}
}
\\

For the singly-charged sector, S-matrix is $8\times8$ matrix with the following two-particle states as rows and columns:
\\

\begin{eqnarray*}
1 \equiv |h_1 w_1^+>,~
2 \equiv|h_1 w_2^+>,
~3\equiv|z_1 w_1^+>,
~4\equiv|z_2 w_2^+>,\nn
\end{eqnarray*}
\begin{eqnarray*}
5\equiv|h_1 w_2^+>,~
6\equiv|h_2 w_1^+>,
~7\equiv|z_1 w_2^+>,
~8\equiv|z_2 w_1^+>.\nn
\end{eqnarray*}
\\
The elements of matrix are given by:
\\
\begin{eqnarray*}
\mathcal{M}^{C}_{1,1}&=& \frac{2 C_{H2} t}{f^2}+2 \left(\frac{9 c_{111} v_1^2}{2 f^2}+\frac{c_{112} v_2^2}{2 f^2}+\frac{c_{(1221)1} v_2^2}{4 f^2}+\frac{c_{(1212)1} v_2^2}{2 f^2}+\lambda _1+\lambda _3\right),\nn\\
\end{eqnarray*}
\begin{eqnarray*}
\mathcal{M}^{C}_{1,2}&=& \frac{2 C_{H12} t}{f^2}+2 \left(\frac{9 c_{111} v_1^2}{2 f^2}+\frac{c_{112} v_2^2}{2 f^2}+\frac{c_{(1221)1} v_2^2}{4 f^2}+\frac{c_{(1212)1} v_2^2}{2 f^2}+\lambda _1+\lambda _3\right),\nn\\
\mathcal{M}^{C}_{1,3}&=&0,\nn\\
\mathcal{M}^{C}_{1,4}&=&\frac{3 i c_{(1221)1} v_1^2}{4 f^2}-\frac{3 i c_{(1212)1} v_1^2}{2 f^2}+\frac{i c_{(1212)1} v_2^2}{4 f^2}-\frac{i c_{(1212)2} v_2^2}{2 f^2}+\frac{i \lambda _6}{2}-\frac{i \lambda _4}{2},\nn\
\mathcal{M}^{C}_{1,5}&=&\frac{3 c_{(1221)1} v_1 v_2}{4 f^2}+\frac{3 c_{(1212)1} v_1 v_2}{2 f^2},\nn\\
\mathcal{M}^{C}_{1,6}&=&
\frac{3 c_{(1221)1} v_1^2}{4 f^2}+\frac{3 c_{(1212)1} v_1^2}{2 f^2}+\frac{3 c_{(1212)1} v_2^2}{4 f^2}+\frac{3 c_{(1212)2} v_2^2}{2 f^2}+\frac{\lambda _5}{2}-\frac{\lambda _4}{2},\nn\\
\mathcal{M}^{C}_{1,7}&=&\frac{i c_{(1221)1} v_1^2}{4 f^2}-\frac{i c_{(1212)1} v_1^2}{2 f^2}+\frac{3 i c_{(1212)1} v_2^2}{4 f^2}-\frac{3 i c_{(1212)2} v_2^2}{2 f^2}+\frac{i \lambda _6}{2}-\frac{i \lambda _4}{2},\nn\\
\mathcal{M}^{C}_{1,8}&=&\frac{3 i c_{(1212)1} v_1^2}{2 f^2}-\frac{3 i c_{(1221)1} v_1^2}{4 f^2}+\frac{i c_{(1212)2} v_2^2}{2 f^2}-\frac{i c_{(1212)1} v_2^2}{4 f^2}+\frac{i \lambda _4}{2}-\frac{i \lambda _6}{2},\nn\\
\mathcal{M}^{C}_{2,2}&=&\frac{2 C_{H2} t}{f^2}2 +\left(\frac{c_{122} v_1^2}{2 f^2}+\frac{c_{(1212)1} v_1^2}{4 f^2}+\frac{c_{(1212)2} v_1^2}{2 f^2}+\frac{9 c_{222} v_2^2}{2 f^2}+\lambda _2+\lambda _3\right),\nn\
\mathcal{M}^{C}_{2,3}&=&\frac{i c_{(1221)1} v_1^2}{4 f^2}-\frac{i c_{(1212)1} v_1^2}{2 f^2}+\frac{3 i c_{(1212)1} v_2^2}{4 f^2}-\frac{3 i c_{(1212)2} v_2^2}{2 f^2}+\frac{i \lambda _6}{2}-\frac{i \lambda _4}{2},\nn
\mathcal{M}^{C}_{2,4}&=&0,\nn\\
\mathcal{M}^{C}_{2,5}&=&\frac{3 c_{112} v_1^2}{2 f^2}+\frac{c_{122} v_2^2}{2 f^2}+\frac{c_{(1212)1} v_2^2}{4 f^2}+\frac{c_{(1212)2} v_2^2}{2 f^2}+\lambda _3+\frac{\lambda _4}{2},\nn\\
\mathcal{M}^{C}_{2.6}&=&\frac{3 c_{(1212)1} v_1 v_2}{4 f^2}+\frac{3 c_{(1212)2} v_1 v_2}{2 f^2},\nn\\
\mathcal{M}^{C}_{2,7}&=&0,\nn\\
\mathcal{M}^{C}_{2,8}&=&\frac{i c_{(1212)1} v_1 v_2}{2 f^2}-\frac{i c_{(1212)2} v_1 v_2}{f^2},\nn\\
\mathcal{M}^{C}_{3,3}&=&\frac{2 C_{H1} t}{f^2}+2 \left(\frac{3 c_{111} v_1^2}{2 f^2}+\frac{c_{112} v_2^2}{2 f^2}+\frac{c_{(1221)1} v_2^2}{4 f^2}-\frac{c_{(1212)1} v_2^2}{2 f^2}+\lambda _1+\lambda _3\right),\nn\\
\mathcal{M}^{C}_{3,4}&=&\frac{2 C_{H12} t}{f^2}+\frac{c_{(1221)1} v_1^2}{4 f^2}+\frac{c_{(1212)1} v_1^2}{2 f^2}+\frac{c_{(1212)1} v_2^2}{4 f^2}+\frac{c_{(1212)2} v_2^2}{2 f^2}+\frac{\lambda _5}{2}-\frac{\lambda _4}{2},\nn\\
\mathcal{M}^{C}_{3,5}&=&\frac{3 i c_{(1212)1} v_1^2}{2 f^2}-\frac{3 i c_{(1221)1} v_1^2}{4 f^2}+\frac{i c_{(1212)2} v_2^2}{2 f^2}-\frac{i c_{(1212)1} v_2^2}{4 f^2}+\frac{i \lambda _4}{2}-\frac{i \lambda _6}{2},\nn\\
\end{eqnarray*}
\begin{eqnarray*}
\mathcal{M}^{C}_{3,6}&=&
\mathcal{M}^{C}_{3,7}=0,\nn\\
\mathcal{M}^{C}_{3,8}&=&
\frac{2 c_{(1212)1} v_1 v_2}{f^2},\nn\
\mathcal{M}^{C}_{4,4}&=&\frac{2 C_{H1} t}{f^2}+2 \left(\frac{c_{122} v_1^2}{2 f^2}+\frac{c_{(1212)1} v_1^2}{4 f^2}-\frac{c_{(1212)2} v_1^2}{2 f^2}+\frac{3 c_{222} v_2^2}{2 f^2}+\lambda _2+\lambda _3\right),\nn\\
\mathcal{M}^{C}_{4,5}&=&
0,\nn\\
\mathcal{M}^{C}_{4,6}&=&\frac{i c_{(1212)1} v_1 v_2}{2 f^2}-\frac{i c_{(1212)2} v_1 v_2}{f^2},\nn\\
\mathcal{M}^{C}_{4,7}&=&\frac{2 c_{(1212)2} v_1 v_2}{f^2},\nn\\
\mathcal{M}^{C}_{4,8}&=&\frac{c_{(1212)1} v_1 v_2}{4 f^2}+\frac{c_{(1212)2} v_1 v_2}{2 f^2},\nn\\
\mathcal{M}^{C}_{5,5}&=& \frac{2 C_{H1H2} t}{f^2}+\frac{3 c_{112} v_1^2}{2 f^2}+\frac{c_{122} v_2^2}{2 f^2}+\frac{c_{(1212)1} v_2^2}{4 f^2}+\frac{c_{(1212)2} v_2^2}{2 f^2}+\lambda _3+\frac{\lambda _4}{2},\nn\\
\mathcal{M}^{C}_{5,6}&=&\frac{2 C_{H12} t}{f^2}+\frac{3 c_{(1221)1} v_1^2}{4 f^2}+\frac{3 c_{(1212)1} v_1^2}{2 f^2}+\frac{3 c_{(1212)1} v_2^2}{4 f^2}+\frac{3 c_{(1212)2} v_2^2}{2 f^2}+\frac{\lambda _5}{2}-\frac{\lambda _4}{2},\nn\\
\mathcal{M}^{C}_{5,7}&=&0,\nn\\
\mathcal{M}^{C}_{5,8}&=&
\frac{3 i c_{(1212)1} v_1^2}{2 f^2}-\frac{3 i c_{(1221)1} v_1^2}{4 f^2}+\frac{i c_{(1212)2} v_2^2}{2 f^2}-\frac{i c_{(1212)1} v_2^2}{4 f^2}+\frac{i \lambda _4}{2}-\frac{i \lambda _6}{2},\nn\\
\mathcal{M}^{C}_{6,6}&=&\frac{2 C_{H1H2} t}{f^2}+\frac{c_{112} v_1^2}{2 f^2}+\frac{c_{(1221)1} v_1^2}{4 f^2}+\frac{c_{(1212)1} v_1^2}{2 f^2}+\frac{3 c_{122} v_2^2}{2 f^2}+\lambda _3+\frac{\lambda _4}{2},\nn\\
\mathcal{M}^{C}_{6,7}&=&\frac{i c_{(1212)1} v_1^2}{2 f^2}-\frac{i c_{(1221)1} v_1^2}{4 f^2}+\frac{3 i c_{(1212)2} v_2^2}{2 f^2}-\frac{3 i c_{(1212)1} v_2^2}{4 f^2}+\frac{i \lambda _4}{2}-\frac{i \lambda _6}{2},\nn\\
\mathcal{M}^{C}_{6,8}&=&0,\nn\\
\mathcal{M}^{C}_{7,7}&=&\frac{2 C_{H1H2} t}{f^2}+\frac{c_{112} v_1^2}{2 f^2}+\frac{c_{122} v_2^2}{2 f^2}+\frac{c_{(1212)1} v_2^2}{4 f^2}-\frac{c_{(1212)2} v_2^2}{2 f^2}+\lambda _3+\frac{\lambda _4}{2},\nn\\
\mathcal{M}^{C}_{7,8}&=&\frac{2 C_{H12} t}{f^2}+\frac{c_{(1221)1} v_1^2}{4 f^2}+\frac{c_{(1212)1} v_1^2}{2 f^2}+\frac{c_{(1212)1} v_2^2}{4 f^2}+\frac{c_{(1212)2} v_2^2}{2 f^2}+\frac{\lambda _5}{2}-\frac{
 _4}{2},\nn\\
\mathcal{M}^{C}_{8,8}&=&\frac{2 C_{H1H2} t}{f^2}+\frac{c_{112} v_1^2}{2 f^2}+\frac{c_{(1221)1} v_1^2}{4 f^2}-\frac{c_{(1212)1} v_1^2}{2 f^2}+\frac{c_{122} v_2^2}{2 f^2}+\lambda _3+\frac{\lambda _4}{2} .\nn\
\end{eqnarray*}

\end{appendices}